\documentclass[iop,numberedappendix,twocolumn]{emulateapj}  
\usepackage[]{natbib, graphicx, hyperref}
\usepackage{CJK}
\usepackage{ulem}

\shorttitle{SMT CO (2-1) Observations of Nearby Star-Forming Galaxies}
\shortauthors{Jiang et al.}

\begin{document}
\begin{CJK*}{UTF8}{gbsn}
\title{SMT CO (2-1) Observations of Nearby Star-Forming Galaxies}

\author{
Xue-Jian Jiang (蒋雪健)\altaffilmark{1,2,3,4},
Zhong Wang\altaffilmark{2},
Qiusheng Gu\altaffilmark{1,3,4},
Junzhi Wang\altaffilmark{5} and 
Zhi-Yu Zhang\altaffilmark{6,7}
}

\altaffiltext{1}{School of Astronomy and Space Science, Nanjing University,
Nanjing 210093, China}
\altaffiltext{2}{Harvard-Smithsonian Center for Astrophysics, MS 66, 60 Garden
St., Cambridge, MA 02138, United States}
\altaffiltext{3}{Key Laboratory of Modern Astronomy and Astrophysics, Nanjing
University, Ministry of Education, Nanjing 210093, China}
\altaffiltext{4}{Collaborative Innovation Center of Modern Astronomy and Space
Exploration, Nanjing 210093, China}
\altaffiltext{5}{Shanghai Astronomical Observatory, Chinese Academy of
Sciences, 80 Nandan Road, Shanghai 200030, China}
\altaffiltext{6}{The UK Astronomy Technology Centre, 
Royal Observatory Edinburgh, Blackford Hill, Edinburgh, EH9 3HJ, United Kindom}
\altaffiltext{7}{ESO, Karl Schwarzschild Strasse 2, D-85748 Garching, Munich,
Germany}

\email{Email: xjjiang@nju.edu.cn}







\begin{abstract} 
We present CO $J$=2-1 observations towards 32 nearby gas-rich star-forming
galaxies selected from the ALFALFA and WISE catalogs, using the Sub-millimeter
Telescope\footnote{The SMT is operated by the Arizona Radio Observatory (ARO),
Steward Observatory, University of Arizona.}. Our sample is selected to be
dominated by intermediate-$M_{\rm *}$ galaxies. The scaling-relations between
molecular gas, atomic gas and galactic properties (stellar mass, NUV$- r$ and
{\it WISE} color W3$-$W2) are examined and discussed. Our results show that (1). In
the galaxies with stellar mass $M_{\rm *}$ $\leqslant 10^{10}$ $M_{\odot}$, H\,{\sc i} fraction
($f_{\rm HI}$ $\equiv$ $M_{\rm H\,{\sc I}}$/$M_{\rm *}$) is significantly higher than that of more massive
galaxies, while H$_2$ gas fraction ($f_{\rm H_2}$ $\equiv$ $M_{\rm H_2}$/$M_{\rm *}$) remain nearly
unchanged. (2).  Comparing with $f_{\rm H_2}$, $f_{\rm HI}$ correlates better with both $M_{\rm *}$ and
NUV$- r$. (3).  A new parameter, {\it WISE} color W3$-$W2 (12\,$\mu$m$-$4.6\,$\mu$m) is
introduced, which is similar to NUV$- r$ in tracing star formation activity,
and we find that W3$-$W2 has a tighter anti-correlation with log $f_{\rm H_2}$ than the
anti-correlation of (NUV$- r$) - $f_{\rm HI}$, (NUV$- r$) - $f_{\rm H_2}$ and (W3$-$W2) - $f_{\rm HI}$.
This indicates that W3$-$W2 can trace the H$_2$ fraction in galaxies. For gas
ratio $M_{\rm H_2}$/$M_{\rm H\,{\sc I}}$, only in the intermediate-$M_{\rm *}$ galaxies it appears to depend on
$M_{\rm *}$ and NUV$- r$. We find a tight correlation between the molecular gas mass
$M_{\rm H_2}$ and 12\,$\mu$m  (W3) luminosities ($L_{\rm 12\,\mu m}$), and the slope is close to unity
(1.03 $\pm$ 0.06) for the SMT sample. This correlation may reflect that the
cold gas and dust are well mixed on global galactic scale. Using the all-sky
12\,$\mu$m (W3) data available in {\it WISE}, this correlation can be used to
estimate CO flux for molecular gas observations and can even predict H$_2$ mass
for star-forming galaxies.  
\end{abstract}


\section{Introduction} 
\setcounter{footnote}{0}

In all star forming systems from Galactic molecular clouds to high
redshift galaxies, cold atomic and molecular gases (H\,{\sc i} and H$_2$) are the raw
material that form stars and drives the evolution of galaxies.  Understanding
of the relationships between cold gas and global properties of galaxies, such
as star formation activities, has been greatly improved thanks to the recent
multi-wavelength advances. For example, galaxies show strongly bimodal
distributions in their integrated colors, which indicates two modes in the
galaxy evolution histories, that young galaxies with active star-formation in
the ``blue cloud'', appeared to rapidly evolve into the ``red sequence''
populated by quiescent galaxies with little star-formation
\citep{Kennicutt:2012}. This transition must be preceded by the transformation
from H\,{\sc i} to H$_2$, the assembly of molecular clouds, and star-formation therein
\citep{Krumholz:2013, Schruba:2011}.

Star forming galaxies were found to fall on a main sequence (MS) in the
relation between stellar mass and star formation rate \citep{Daddi:2007}, and
recent studies found that the scatter in the MS is remarkably small at
different redshifts, implying that the majority of star forming galaxies are
`normal' galaxies lying along the relation \citep{Guo:2013, Rodighiero:2011}.
Particularly, intermediate-mass (or low mass) galaxies ($M_{\rm *} < 10^{10}
M_{\odot}$) are still not well understood. Their number dominates the galaxy
populations, many of them are cold ISM dominated and have lower metallicities
than solar value. As a result their star formation properties are distinct from
massive galaxies. Low mass galaxies were found to be more gas rich and
inefficient in transforming their gas into stars \citep{Blanton:2009}, but they
may dominate the present star-forming galaxy populations rather than massive
galaxies (downsizing effect), and may help understand high-$z$ galaxies since the
physical conditions in low mass systems such as high gas ratio and low
metallicity resemble those at the early universe.

Recent studies have found a steep decline of cosmic star formation rate density
between $z = 1$ and 0, and peaks at $z \sim 1-2 $
\citep{Bouwens:2011,Carilli:2013}, and this evolution appears to be strongly
regulated by the evolution of cold gas content, of which the most important
parameters are gas fraction $f_{\rm gas}$ ($M_{\rm H_2}$/$M_{\rm *}$ and $M_{\rm H\,{\sc I}}$/$M_{\rm *}$, sometimes
$M_{\rm gas}$/($M_{\rm *}$ + $M_{\rm gas}$)) and gas phase ratio $r_{\rm gas}$ ($M_{\rm H_2}$/$M_{\rm H\,{\sc I}}$).
$f_{\rm gas}$ and $r_{\rm gas}$ have been found to be significantly increased
toward high redshift, and the pronounced evolution of $f_{\rm gas}$ seems to
resemble the evolution of SFR \citep{Combes:2013}, implies its important role
in determining the star formation history. Moreover, 
\citet{Tacconi:2013} shows a trend of increasing $f_{\rm gas}$ with decreasing
stellar mass \citep[see also][]{Saintonge:2011}, implying that the cosmic gas
density could be higher than current derived values \citep{Carilli:2013}.
Although representing main-sequence star-forming galaxies, the sample by
\citet{Tacconi:2013} were limited to the high-mass end of galaxies distribution,
and studies on low- to intermediate-stellar mass galaxies would provide crucial
constraint on the cosmic gas content as well as star-formation history.  In
this context, the study of nearby galaxies of intermediate-stellar mass can
provide a benchmark of high-$z$ studies.  In existing CO surveys, galaxies
beyond our immediate vicinity are almost entirely dominated by massive
systems \citep[COLD GASS,][]{Saintonge:2011} and/or Luminous Infrared Galaxies
\citep[LIRGs,][]{Genzel:2010}, while surveys of intermediate--mass galaxies
were limited by distance \citep[HERACLES,][]{Leroy:2009} or environment
\citep[AMIGA,][]{Lisenfeld:2011}.  This stands in sharp contrast to surveys of
other wavelengths (e.g., SDSS in optical or GALEX in UV), where homogeneous
samples of ordinary (and many low--mass) galaxies out to hundreds of Mpc
distance are available.

Atomic gas (neutral hydrogen) is traced by H\,{\sc i} 21-cm hyperfine transition line
\citep{Haynes:2011} and molecular gas (mainly H$_2$) mass can be traced by carbon
monoxide, typically $^{12}$CO(J=1$\rightarrow$0) \citep{Young:1991,
Saintonge:2011, Lisenfeld:2011} and $^{12}$CO(J=2$\rightarrow$1) \citep{Leroy:2013}. Surveys towards
very nearby galaxies such as SINGS \citep{Kennicutt:2003} and THINGS
\citep{Walter:2008} have conducted detailed analysis for spiral disk of
galaxies, while there are also surveys with larger sample of greater distances
\citep[$>$ 100 Mpc, e.g., the COLD GASS,][]{Saintonge:2011, Saintonge:2011b},
that mainly focus on the global scaling relations.  Much recent effort has been
devoted to study of the correlation between molecular gas tracers (CO, HCN,
HCO$^+$ and CS etc.) and star formation in nearby galaxies, both individually
and as a function of Hubble type \citep{Zhang:2014, Gao:2004, Calzetti:2007,
Kennicutt:2009, Bigiel:2011}.  On global scales ($\gtrsim$ 1 kpc) H\,{\sc i} and
H$_2$ appear to depend on other galactic physical properties, and such
scaling-relations can help better understand the role of cold gas in the
histories of galaxy evolution and star formation.

The Arecibo ALFALFA survey \citep[$\alpha$.40,][]{Giovanelli:2005} scanned a
large area of the sky and aimed to study the H\,{\sc i} content of normal galaxies out
to $\sim$ 250 Mpc \citep{Haynes:2011}. It so far covers $\sim$ 3,000 square
degrees of the sky and contains more than 15,000 galaxies, and is characterized
by a maximum overlap with the coverage of both the SDSS and GALEX. Comparing
with optical-selected sample, this H\,{\sc i}-selected sample represents a less
evolved galaxy population, in which a transition in star formation properties
is found at around $M_{*} \sim 10^{9.5} M_{\odot}$.  Below this mass, galactic
star formation is found to be strongly regulated by H\,{\sc i} mass rather than
stellar mass \citep{Huang:2012}. Thus a survey of molecular gas in the
intermediate-mass galaxies selected from the ALFALFA sample would provide a key
missing link between star formation and global gas content (H$_2$ + H\,{\sc i}) in these
galaxies.  

In this paper, we describe our SMT sample in Section \ref{sec:sample}.  The SMT
observation and results are presented in Section \ref{sec:obs}.  We present our
calculations of molecular gas mass for the SMT sample in Section \ref{sec:mh2},
as well as the stellar mass calculations for both the SMT sample and the AMIGA
sample. The main results are presented in section \ref{sec:results}. In section
\ref{sec:sample-comparison} the different samples are compared,  in sections
\ref{sec:fraction} and \ref{sec:ratio} we discuss the scaling relations of gas
fraction ($M_{\rm gas}$/$M_{\rm *}$) and gas ratio ($M_{\rm H_2}$/$M_{\rm H\,{\sc I}}$), and  in Section
\ref{sec:correlation} the correlation between molecular gas and {\it WISE} infrared
properties is discussed.  Throughout this work, we assume the cosmological
parameters $(\Omega_{\rm m} = 0.3, \Omega_{\Lambda} = 0.7, H_{0} = 70)$ which
are consistent with the \textit{Wilkinson Microwave Anisotropy Probe} Third
Year results in combination with other data \citep{Spergel:2007}.

\section{The Sample} \label{sec:sample} 
\subsection{The SMT sample} To obtain a representative sample of nearby
galaxies selected based on gas content and to explore their infrared
properties, we begin from the most up-to-date ALFALFA survey catalog and look
for counterparts in the {\it WISE} all-sky catalog to construct the parent sample.
Comparing with optical-selected samples, the ALFLAFA is an H\,{\sc i}-selected sample
thus represents a less evolved galaxy population.

The Wide-field Infrared Survey Explorer \citep[{\it WISE},][]{Wright:2010}
Catalog provide all-sky image data in four infrared bands (3.4, 4.6, 12 and
22\,$\mu$m), with an angular resolution of 6.1\arcsec, 6.4\arcsec, 6.5\arcsec~
and 12.0\arcsec, respectively. {\it WISE} achieved 5$\sigma$ point source sensitivities better
than 0.08, 0.11, 1 and 6 mJy, respectively, in unconfused regions on the
ecliptic in the four bands.

First, we cross--matched the ALFALFA $\alpha$.40 catalog with the {\it WISE} all-sky
catalog, and selected galaxies based on the following criteria: (1) S/N
$\geqslant$ 5 for the W1 (3.4\,$\mu$m), W2 (4.6\,$\mu$m) and W3 (12\,$\mu$m) bands of {\it WISE}, and S/N
$\geqslant$ 3 for W4 (22\,$\mu$m).  Because W4 is much less sensitive than other bands,
and we are not using W4 in our analysis, so we lower the requirement for the
W4 band; (2) An object is defined as being the same one in
both catalogs if the difference between its coordinates in {\it WISE} catalog and
ALFALFA catalog is $\leqslant$ 3\arcsec. We obtained 5434 ALFALFA--{\it WISE}
counterparts as a parent sample.

Second, based on this parent sample, we limited the distance
of the sample to be in the range of 20--165 Mpc, and required that galaxies 
are covered by the GALEX MIS (Medium Image Survey, exposure time $>$ 1500s). 

Third, we selected infrared bright (W3 $\leqslant$ 10 and W4 $\leqslant$ 8) galaxies to
assure a high detection probability within realistic integration time.  The
stellar masses of the sample, which were obtained from the MPA-JHU
catalog\footnote{\url{http://www.mpa-garching.mpg.de/SDSS/}}, 
are then selected to be distributed in the range of
10$^8 - 10^{11}$ $M_{\odot}$. Therefore we got 115 galaxies as candidates for the
observation. For these galaxies, we visually checked the optical images of
these galaxies from SDSS to exclude those obvious interacting or merger
systems, to assure that the SMT telescope only obtain CO emission from the
target galaxies. we used the W3 magnitude and H\,{\sc i} linewidth to estimate the CO
flux of the sample, and observed those relatively strong sources. 

30 galaxies were selected from this ALFALFA--{\it WISE}
parent sample, and other two LIRGs (Luminous Infrared Bright Galaxies) were
added for comparison. Hence there are 32 galaxies in the final SMT sample.
Although the selection might biases the sample to gas-rich and/or infrared
bright galaxies, these galaxies still span a range in a set of more fundamental
parameters, such as $M_{\rm H_2}$, $M_{\rm H\,{\sc I}}$ and $M_{\rm *}$, and most of them have stellar masses $<
10^{10}$$M_{\odot}$, so that this sample can still fulfill our goal that focus on
intermediate-$M_{\rm *}$ galaxies.  We list the general information of the SMT sample
in Table \ref{tab:sample}.

\subsection{Literature CO 1-0 data} Since there were a few molecular gas
surveys conducted recently, we combined our SMT sample which was selected from
the ALFLAFA-{\it WISE} parent sample, with these surveys, so that we can
compare the properties and/or relations of different galaxies.  

COLD GASS \citep[CO Legacy Database for the GASS, ]
[]{Saintonge:2011, Catinella:2012} is a molecular gas survey in nearby massive
($M_{\rm *}$ $\geqslant$ 10$^{10}$$M_{\odot}$) galaxies. By measuring CO 1-0 with the IRAM 30-m
telescope, they have provided $M_{\rm H_2}$ for 366 galaxies, 194 of which were treated
as secure CO 1-0 detection (S/N $\geqslant$ 5).
 
AMIGA (Analysis of the interstellar Medium of Isolated GAlaxies) is another
galaxy survey in nearby isolated galaxies, which aimed to address the
environmental effects on cold gas content of galaxies. \cite{Lisenfeld:2011}
has provided CO 1-0, $M_{\rm H_2}$ as well as $M_{\rm H\,{\sc I}}$ data for 273 AMIGA galaxies, of which
180 were from observations from IRAM 30-m, FCRAO 14-m (Five College Radio
Astronomical Observatory) and the rest from other literature.

We searched in the {\it WISE} all-sky catalog for counterparts of the COLD GASS and
AMIGA-CO sample. There are {\it WISE} counterparts for 349 COLD GASS galaxies, and
179 AMIGA-CO galaxies, respectively. Among these counterparts, 183 COLD GASS
sources have S/N $\geqslant$ 5 in the W1, W2, W3 bands as well as CO detection,
and 86 AMIGA-CO sources satisfied the same criteria. In our discussion of the
relationship between {\it WISE} infrared emission and molecular gas, we only use
these high S/N sample. When exploring the relation between CO emission and
{\it WISE} infrared emission, we found a tight correlation between the CO and
12\,$\mu$m emission for both COLD GASS and AMIGA survey (we will discuss this
relation in detail in Sec. \ref{sec:correlation}).  With these combined
sample, we can therefore investigate the scaling-relations in
intermediate-$M_{\rm *}$ galaxies, and avoid the possible environmental effect induced
in the AMIGA-CO sample.


\subsection{SDSS and {\it GALEX} data} In order to construct a multi-wavelength
dataset for comprehensive analysis and comparison with the COLD GASS and
AMIGA--CO, We have collected SDSS and {\it GALEX} data for both the SMT sample and
the AMIGA--CO sample.

The newest data release of Sloan Digital Sky Survey, SDSS--III DR9 (hereafter
DR9), have provided more available images and photometric measurements in
\textit{ugriz} bands. We run query with our SMT sample and the AMIGA--CO sample
in DR9, and obtained photometric magnitudes ({\it ugriz}, extinction
corrected) and morphological parameters (R90 and R50) for 30 SMT galaxies and
186 AMIGA--CO galaxies.  Since extended sources are probably ``deblended" in
the SDSS photometry process \citep{Huang:2012}, we check the SDSS flags for
each galaxy to make sure that they are the ``brightest child" photometric
object extracted from the galaxy.

However, one SMT galaxy, AGCNr 4584, has an abnormal {\it z}-band magnitudes in
DR9 ({\it z}$_{\rm dr9}$ = 22.9$^{\rm mag}$), which significantly deviates from
its value in DR7 ({\it z}$_{\rm dr7}$ = 13.1$^{\rm mag}$) as well as other
galaxies' {\it z} -band magnitudes, which are all brighter than 16$^{\rm mag}$.
After inspecting the SDSS images of AGCNr\,4584, we found that the photometric
position of this galaxies has been shifted in DR9 (and DR8) but is not in the
galactic nucleus, which may be the reason of the abnormal photometry.  Thus for
AGCNr 4584, we adopt its data from DR7 instead of DR9.  

We also searched in the {\it GALEX} GR6/7 catalog and found {\it GALEX} counterparts for
27 SMT galaxies and 223 AMIGA-CO galaxies. For the AMIGA-CO sample, 9\% SDSS
counterparts and 28\% {\it GALEX} counterparts have angular separation larger than
3\arcsec, but we did visual inspection and assure that they are the genuine
counterparts and are not contaminated by nearby galaxies.  In fact, since the
AMIGA sample are all isolated galaxies, mismatch and/or contamination is not
likely for nearly all the AMIGA-CO galaxies. The NUV magnitudes is corrected
for Galactic extinction following \citet{Wyder:2007}.


\begin{table} \begin{center}
\caption{General data of the SMT sample}
  \scriptsize
\begin{tabular}{cccrrr}
\hline
\hline
galaxy &\multicolumn{2}{c}{RA  (J2000)   Dec}& Distance & log $M_{\rm HI}$\tablenotemark{a}  & $r_{50}$\tablenotemark{b} \\
&[h~\phn m~\phn s] & [\phn \degr~\phn \arcmin ~\phn \arcsec] & [Mpc] & [$M_{\odot}$]  & [\arcsec] \\
\hline
AGC\,230    & 00:23:59.4 & +15:46:14 & 74.8 & 10.2  &11.5 \\
AGC\,233    & 00:24:42.8 & +14:49:28 & 74.3 & 9.7   &6.5 \\
AGC\,463    & 00:43:32.4 & +14:20:33 & 62.1 & 9.7   &16.8\\
AGC\,978    & 01:25:13.2 & +14:52:20 & 86.6 & 9.9   &11.6\\
AGC\,629    & 02:08:25.0 & +14:20:57 & 60.6 & 10.0  &9.4  \\
AGC\,456    & 08:32:03.5 & +24:00:37 & 79.8 & 10.0  &14.6 \\
AGC\,473    & 08:33:56.7 & +26:58:20 & 51.9 & 9.7   &8.2 \\
AGC\,584    & 08:45:59.5 & +12:37:11 & 61.3 & 9.9   &16.3\\
AGC\,760    & 09:04:54.2 & +25:00:12 & 44.7 & 9.1   &12.2\\
AGC\,823    & 09:10:41.8 & +07:12:23 & 23.8 & 9.1   &7.6 \\
AGC\,122    & 14:15:23.8 & +04:24:28 & 86.2 & 9.8   &7.5 \\
AGC\,172    & 14:20:04.4 & +03:59:32 & 27.8 & 9.1   &15.6\\
AGC\,9346   & 14:31:57.4 & +06:15:00 & 36.3 & 9.8   &12.1\\
AGC\,9696   & 15:05:30.5 & +08:31:26 & 121.5 &10.0  &6.7  \\
AGC\,10384  & 16:26:46.6 & +11:34:49 & 74.4 & 10.1  &7.4  \\
AGC\,12481  & 23:17:36.6 & +14:00:02 & 63.4 & 9.8   &6.5 \\
AGC\,12586  & 23:24:49.3 & +15:16:31 & 61.2 & 9.8   &11.1\\
AGC\,101985 & 00:20:48.6 & +14:13:28 & 75.5 & 9.0   &2.0 \\
AGC\,110202 & 01:19:32.9 & +14:52:19 & 56.4 & 9.1   &5.4 \\
AGC\,171860 & 07:49:56.0 & +26:02:13 & 70.8 & 9.2   &5.5 \\
AGC\,180293 & 08:28:22.3 & +25:07:29 & 32.8 & 8.4   &9.2 \\
AGC\,180931 & 08:07:10.3 & +24:23:21 & 157.8 & 9.2  &7.2  \\
AGC\,180945 & 08:09:45.5 & +25:52:50 & 110.7 & 9.8  &5.5  \\
AGC\,240580 & 14:40:22.7 & +09:28:34 & 124.0 & 9.2  &7.0  \\
AGC\,241492 & 14:20:42.2 & +05:29:08 & 120.7 &10.4  &4.4  \\
AGC\,241716 & 14:55:21.0 & +05:17:52 & 95.2 & 9.7   &6.4 \\
AGC\,242440 & 14:08:46.8 & +04:54:35 & 76.9 & 9.6   &1.9 \\
AGC\,250019 & 15:03:30.7 & +10:49:47 & 147.9 & 9.5  &4.1  \\
AGC\,330149 & 23:14:59.5 & +14:59:19 & 165.0 & 9.8  &9.6  \\
AGC\,721352 & 09:07:08.3 & +25:20:32 & 41.8 & 8.5   &7.8 \\
PGC\,003183 & 00:54:03.6 & +73:05:12 & 68.0 & 9.9   & \ldots\\
UGC\,11898  & 22:04:36.1 & +42:19:38 & 61.9 & 9.0   & \ldots\\
\hline
\end{tabular}\label{tab:sample}
\end{center}
Note: \tablenotemark{a} $M_{\rm H\,{\sc I}}$ are from the ALFALFA $\alpha$.40 catalog.\\
\tablenotemark{b} Half-light radius $r_{50}$ in unit of arcsec, 
containing 50\% of the $r$ band flux.
\end{table}

\begin{table}
\begin{center}
\caption{Estimated data of the SMT sample}
  \scriptsize
\begin{tabular}{ccrcrr}
\hline
\hline
galaxy & RMS\tablenotemark{a}  & FWHM & $I_{\rm CO}$ & log $M_{\rm H_2}$ & log $M_{\rm *}$ \\
 & [mK] & [km s$^{-1}$] & [K km s$^{-1}$] & [$M_{\odot}$] & [$M_{\odot}$] \\
\hline
AGC\,230    & 1.6 & 104.2 & 1.20$\pm$0.09 & 9.08 & 9.88 \\
AGC\,233    & 1.4 & 105.7 & 0.85$\pm$0.10 & 8.92 & 9.76 \\
AGC\,463    & 2.5 & 168.8 & 7.09$\pm$0.18 & 9.69 & 10.42 \\
AGC\,978    & 0.9 & 125.0 & 0.43$\pm$0.05 & 8.79 & 9.95 \\
AGC\,629    & 0.9 & 311.6 & 4.98$\pm$0.07 & 9.53 & 10.33 \\
AGC\,456    & 1.3 & 63.4  & 0.86$\pm$0.06 & 8.96 & 10.15 \\
AGC\,473    & 1.8 & 42.6  & 0.28$\pm$0.04 & 8.09 & 9.37 \\
AGC\,584    & 2.6 & 53.2  & 1.52$\pm$0.08 & 8.96 & 9.56 \\
AGC\,760    & 1.0 & 147.1 & 1.46$\pm$0.07 & 8.66 & 9.72 \\
AGC\,823    & 1.2 & 62.8  & 0.47$\pm$0.06 & 7.53 & 8.28 \\
AGC\,122    & 2.2 & 31.6  & 0.52$\pm$0.06 & 8.79 & 9.49 \\
AGC\,172    & 2.0 & 163.8 & 2.68$\pm$0.16 & 8.46 & 9.66 \\
AGC\,9346   & 2.0 & 187.5 & 1.61$\pm$0.15 & 8.49 & 9.28 \\
AGC\,9696   & 2.7 & 85.2  & 1.86$\pm$0.13 & 9.65 & 10.62 \\
AGC\,10384  & 1.8 & 359.4 & 4.64$\pm$0.20 & 9.60 & 9.98 \\
AGC\,12481  & 3.3 & 57.6  & 2.54$\pm$0.16 & 9.24 & 9.23 \\
AGC\,12586  & 1.6 & 116.0 & 2.85$\pm$0.13 & 9.26 & 10.49 \\
AGC\,101985 & 1.0 & 46.4  & 0.16$\pm$0.04 & 8.19 & 9.06 \\
AGC\,110202 & 1.1 & 63.2  & 0.30$\pm$0.06 & 8.25 & 9.10 \\
AGC\,171860 & 1.1 & 42.1  & $<$ 0.09 & $<$7.89 & 9.00 \\
AGC\,180293 & 1.7 & 42.0  & 0.86$\pm$0.07 & 8.15 & 9.10 \\
AGC\,180931 & 1.3 & 16.0  & 0.11$\pm$0.03 & 8.67 & 10.40 \\
AGC\,180945 & 0.9 & 63.7  & 0.23$\pm$0.04 & 8.66 & 9.13 \\
AGC\,240580 & 1.7 & 106.6 & 0.48$\pm$0.09 & 9.08 & 9.78 \\
AGC\,241492 & 0.8 & 106.5 & 0.52$\pm$0.03 & 9.08 & 9.93 \\
AGC\,241716 & 3.1 & 21.6  & 0.44$\pm$0.07 & 8.81 & 9.64 \\
AGC\,242440 & 2.5 & 32.2  & 0.21$\pm$0.06 & 8.31 & 8.75 \\
AGC\,250019 & 1.2 & 97.1  & 0.31$\pm$0.05 & 9.06 & 9.50 \\
AGC\,330149 & 1.8 & 32.6  & 0.54$\pm$0.05 & 9.44 & 10.37 \\
AGC\,721352 & 1.3 & 63.0  & 1.13$\pm$0.07 & 8.48 & 9.40 \\
PGC\,003183 & 2.5 & 105.5 & 19.96$\pm$0.23 & 10.19 & \ldots \\
UGC\,11898  & 2.5 & 242.6 & 7.79$\pm$0.23 & 9.70 & \ldots \\
\hline
\end{tabular} \label{tab:measurements}
\end{center}
Note: This table list the measurements of CO 2-1 spectra, including noise level,
line width, line integrated intensity, $M_{\rm H_2}$ and $M_{\rm *}$.\\
\tablenotemark{a} RMS noise level of the smoothed spectra.  
\end{table}

\section{SMT Observation and Data Reduction}
\label{sec:obs}

The SMT observations were carried out in December, 2012. We spent 59 hours in
total, including $\sim40\%$ overhead, to observe CO 2-1 ($\nu_{\rm rest}$ =
230.538 GHz) towards 32 galaxies. We used the Forbes Filter Bank system, which
provide 1 MHz per channel and 1024 channels in total, corresponding to a
velocity coverage of 1300 km\,s$^{-1}$ bandwidth for each polarization. During the
observations the typical system temperature was less than 300 K.  Each galaxy
was integrated until CO 2-1 was detected or three hours integration time was
reached. Beam Switch mode with 2$\arcmin$ throw were used and pointing and
focus were checked every 1.5 hours using DR\,21. Saturn was used for flux
calibration. 
The HPBW (half-power beamwidth) of the SMT at this frequency is
about 33$\arcsec$.  

The data were reduced with the \texttt{class} program of the \texttt{gildas}
package\footnote{\url{http://iram.fr/IRAMFR/GILDAS/}}. Poor scans and bad
channels were flagged, and a three degree polynomial baseline was fitted to
subtract the unstable baseline from each spectrum. The root-mean-square (RMS)
noise level of each averaged spectrum of the sample target was obtained at a
velocity resolution of 11 to 21 km\,s$^{-1}$, based on the linewidth of each galaxy. 

The velocity-integrated intensities of CO 2-1 are calculated using
\begin{equation}
I = \int _{\Delta V} T_{\rm mb}{\rm d}v,
\end{equation}

where $T_{\rm mb}$ is the main beam brightness temperature, and $\Delta V$ is
the velocity range used to integrate the intensity. Molecular line intensity in
antenna temperature $T_a^{*}$ is converted to main beam temperature $T_{\rm
mb}$ using $T_{\rm mb} =  \eta_{\rm eff} T_a^{*}$, where the efficiency factor
$\eta_{\rm eff} = F_{\rm eff}/B_{\rm eff} = 77\% $ ($F_{\rm eff}$: forward
efficiency; $B_{\rm eff}$: Beam efficiency) was measured in the flux
calibration. The peak $T_{\rm mb}$ of the sample are in the range between about
3 and 80 mK. The basic measured results of the SMT sample, including their RMS,
linewidth and integrated intensities, are listed in Table
\ref{tab:measurements}.  

The CO 2-1 Spectra of the 32 galaxies obtained by the SMT are shown in
Figure.\,\ref{fig:spec} in the Appendix. Only one galaxy AGCNr\,171860 was a
non-detection, and the 3$\sigma$ upper limits of its CO integrated intensity
and $M_{\rm H_2}$ are derived, with its $\Delta V$ in equation (1) set to the H\,{\sc i}
linewidth ($W_{50}$) from the ALFALFA catalog. Table\,\ref{tab:measurements}
also list the calculated $M_{\rm H_2}$ and $M_{\rm *}$ of the sample, and some technical
details are discussed in the following sections. In Figure.\,\ref{fig:spec} the
radial velocities were normalized to that of their H\,{\sc i} velocities from ALFALFA
($V_{helio}$), and
nearly all the CO 2-1 velocities are consistent with their H\,{\sc i} velocities,
except for a few that have low S/N (AGCNr 978, 101985, 241492, 242440).
It is unlikely that the H$_2$ and H\,{\sc i} are spatially inconsistent, instead we
suggest that such offset between $V_{\rm HI}$ and $V_{\rm CO}$ are at 
least partially resulted from the observational uncertainties.

Figure.\,\ref{fig:spec} also shows that, for some SMT galaxies, especially those
have large line widths, their CO 2-1 spectra exhibit some structures, such as
offset peaks and/or multiple peaks (e.g., AGCNr 463, 1629, 9172, 10384).
Although we do not have spatially-resolved CO images for these galaxies, their
spectra do reflect the clumpy distribution of $M_{\rm H_2}$ in late type galaxies, and
the offset CO peaks suggest that $M_{\rm H_2}$ may locate in circumnuclear regions
and/or spiral arms in the form of giant molecular clouds, which is consistent
with that reveal by imaging surveys \citep[e.g.,][]{Rahman:2012}.

\subsection{Calculation of $M_{\rm H_2}$ and $M_{\rm *}$} \label{sec:mh2}
The total molecular gas of a galaxy is computed using $M_{\rm H_2}$$ = \alpha_{\rm CO}
L'_{\rm CO}$, and the CO line luminosity is calculated following
\cite{Solomon:1997}: 
\begin{equation}
  L'_{\rm CO} = 3.25 \times 10^7 S_{\rm CO} \Delta V \nu_{\rm obs}^{-2} D_L^2 (1+z)^{-3}
\end{equation}
where CO line intensity $S_{\rm CO} \Delta V$ in units of Jy\,km\,s$^{-1}$, observing
frequency $\nu_{\rm obs}$ in GHz, luminosity Distance $D_L$ in Mpc, and $L'_{\rm CO}$ in
K\,km\,s$^{-1}$\,pc$^2$.

We adopt a CO\,2-1/CO\,1-0 line ratio (R21) of 0.7, and a CO(1-0)-to-H$_2$
conversion factor $\alpha_{\rm CO}$ = 4.35 $M_{\odot}$(K\,km\,s$^{-1}$\,pc$^2$)$^{-1}$, equivalent to
$X_{\rm CO} = 2\times 10^{20}$ cm$^{-2}$(K\,km\,s$^{-1})^{-1}$, which are
consistent with that used in COLD GASS and AMIGA-CO, as well as recent studies
of resolved star-forming regions in nearby galaxies \citep{Leroy:2013}. For
more detail about R21 and $\alpha_{\rm CO}$ in nearby galaxies, we refer the readers to
\citet{Leroy:2013} and \citet{Sandstrom:2013}. All the molecular gas masses in
this paper include a factor of 1.36 correcting for the presence of heavy elements (mostly
helium). Although the $M_{\rm H_2}$ could not be directly compared with that derived
from CO 1-0, the HERACLES survey \citep{Leroy:2013, Sandstrom:2013} have
demonstrated that CO 2-1 is able to robustly trace global molecular gas.  

The mean error brought by the line measurements itself is about 10\%, and other
error sources including calibration errors (flux, pointing), missing CO flux
due to the finite dish aperture, and the uncertainty of $\alpha_{\rm CO}$.
\cite{Bolatto:2013} summarized an uncertainty of $\pm$0.3 dex (a factor of 2)
for $\alpha_{\rm CO}$ in the disks of normal, solar metallicity star-forming galaxies.
Therefore the total error on log $M_{\rm H_2}$ is about 0.4 dex. 
We do not adopt aperture correction for this SMT sample, 
since the angular sizes of the galaxies were carefully selected so that
the SMT beam could effectively cover the CO emission region
(their half-light radii are listed in Table \ref{tab:sample}), as CO emitting 
size is generally proportional to the optical size of disk galaxies
\citep{Lisenfeld:2011}.

We calculat the stellar masses with the SED fitting code \texttt{kcorrect}
\citep[version 4.2,][]{Blanton:2007}.  Provided a set of photometric data (in
our case \textit{ugriz} magnitudes), this code return stellar masses fitted
from galaxy templates that based on stellar population synthesis models
\citep{Bruzual:2003} using the \cite{Chabrier:2003} stellar initial mass
function. For consistency we used this method to recalculate $M_{\rm *}$ for those SMT
galaxies and the AMIGA-CO sample that have available SDSS DR9 data, and the
results of the SMT sample is consistent with that provided by the JHU-MPA
catalog. The COLD GASS survey has provided $M_{\rm *}$ data.  $M_{\rm H_2}$, $M_{\rm *}$ of the SMT
sample are listed in Table \ref{tab:measurements}.  The stellar mass for the
186 AMIGA-CO galaxies will be available online at{\it VizieR}.


\section{Results and Discussion} \label{sec:results}

\subsection{Sample Comparison}\label{sec:sample-comparison}

Among all kinds of galactic parameters, stellar mass ($M_{\rm *}$) is the most
fundamental and important one. The history of galaxy evolution is accompanied
by the formation of stars from molecular gas and the cumulation of stars. So $M_{\rm *}$ is
the first parameter being discussed along with H\,{\sc i} and H$_2$ in the following
sections. We use NUV$- r$ as an indicator of star formation activity, as it is
well correlated with specific star formation rate (sSFR).  The third parameter
used in our analysis is {\it WISE} color W3$-$W2, which is similar to NUV$- r$ but
less affected by dust attenuation. 

\begin{figure*}[ht]
  \begin{center}
	\includegraphics[width=18cm, angle=0]{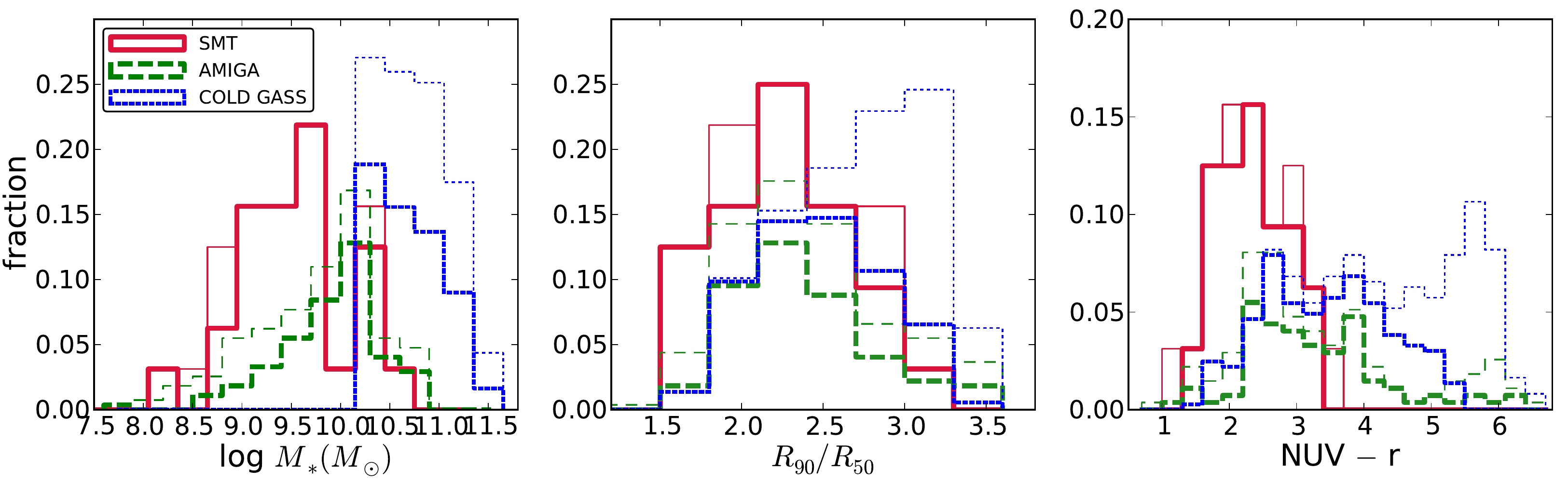}
  \end{center}
  \caption{Normalized distributions of stellar mass $M_{\rm *}$, concentration indices
  $R_{90}/R_{50}$ and NUV$- r$ color of the three samples used in this paper. Blue
  dotted histograms represent the COLD GASS survey of massive galaxies
  ($M_{\rm *}$ $\ge 10^{10}$$M_{\odot}$).  Red solid and green dashed histograms represents
  the SMT sample and the AMIGA-CO, respectively. The histograms of thin lines
  denote the entire sample, while thicker histograms denote only those galaxies
  with secure CO detections (S/N $\ge$ 5). For clarity, in the first panel the
  SMT and COLD GASS are slightly offset with respect to the AMIGA.}  \label{fig:hist1}
\end{figure*}

First of all, we compare our SMT sample with the AMIGA and COLD GASS samples to
explore their difference in physical properties.  Figure.\,\ref{fig:hist1} shows
the distributions of their stellar masses ($M_{\rm *}$), concentration indices
($R_{90}/R_{50}$) and NUV$- r$. Comparing all galaxies from the three samples
and including those undetected in CO, we find a few differences between the
samples. First, the SMT sample is  very similar to the AMIGA especially in
$M_{\rm *}$ and $R_{90}/R_{50}$ distributions, although its NUV$- r$ colors are
slightly bluer than that of AMIGA on average, and AMIGA included a small
portion of galaxies with red NUV$- r$ colors. Second, The COLD GASS galaxies
show distinct properties than the SMT and AMIGA samples, that they are not only
more massive ($M_{\rm *}$ $\ge 10^{10}$$M_{\odot}$), but also tend to be more bulge-dominated
(higher $R_{90}/R_{50}$ values) and redder in NUV$- r$. However, if we only
compare those galaxies with secure CO detections (thicker histograms in
Figure.\,\ref{fig:hist1}), the differences between the three samples become less
prominent, since most of the undetected galaxies are red and budge-dominated.
The NUV$- r$ distribution shows that the COLD GASS contained some red
galaxies, which are probably in accordance with those in the middle panel that
have high $R_{90}/R_{50}$ values. These undetected galaxies should be
early-type massive galaxies that have a predominant stellar bulge, and
quiescent in star formation since they contain little molecular gas.

\begin{figure}[t]
  \begin{center}
	\includegraphics[angle=0,scale=.5]{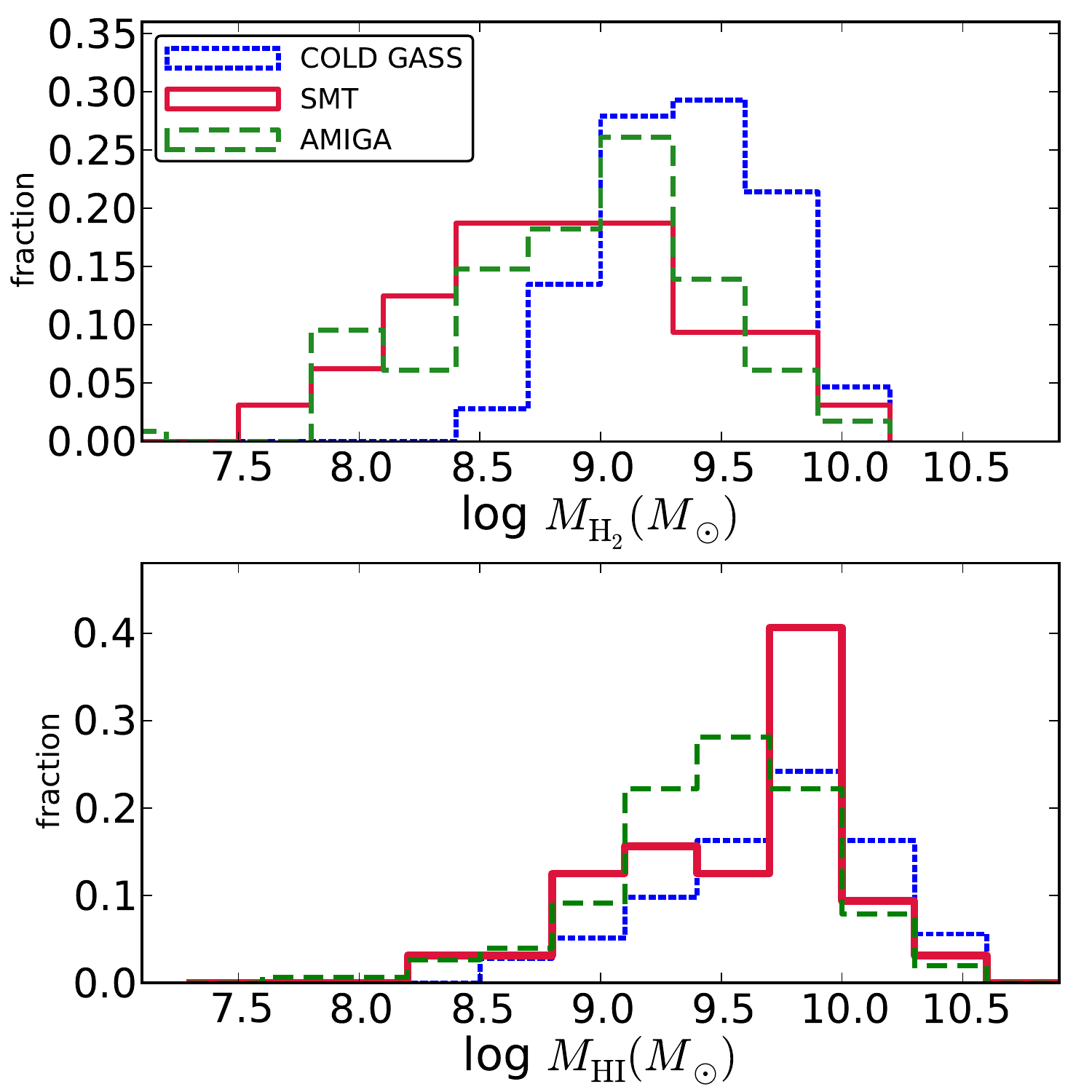}
  \end{center}
  \caption{Normalized distributions of molecular gas masses ($M_{\rm H_2}$) and atomic
  gas masses ($M_{\rm H\,{\sc I}}$) of the three samples used in this paper. Blue histograms
  represent the COLD GASS survey of massive galaxies ($M_{\rm *}$ $\ge 10^{10}$$M_{\odot}$).
  Red solid and green dashed histograms represent our SMT sample and the AMIGA
  sample, respectively. The COLD GASS has narrower $M_{\rm H_2}$  distribution, and the
  three samples are similar in the $M_{\rm H\,{\sc I}}$ distribution.}
  \label{fig:hist2}
\end{figure}

Figure.\,\ref{fig:hist2} shows the normalized distributions of $M_{\rm H_2}$ and $M_{\rm H\,{\sc I}}$ of the
three samples. It is interesting to see that although they share a common
distribution in $M_{\rm H\,{\sc I}}$, their $M_{\rm H_2}$ distributions are different. The highest $M_{\rm H_2}$
in each sample are similarly at $\sim 10^{10.2}$$M_{\odot}$, but the COLD GASS
galaxies have higher mean $M_{\rm H_2}$ and narrower $M_{\rm H_2}$ distribution. Looking into the
samples and we found that, among all the galaxies studied in this paper, only
seven galaxies with $M_{\rm H_2}$ $\le 10^{8.5}$$M_{\odot}$~are massive galaxies ($M_{\rm *}$ $\ge
10^{10}$$M_{\odot}$). So the reason we do not see low $M_{\rm H_2}$ ($\le 10^{8.4}$$M_{\odot}$) in
the COLD GASS sample is because $M_{\rm *}$ is proportional to $M_{\rm H_2}$, and those galaxies
with low $M_{\rm H_2}$ are most likely low $M_{\rm *}$ galaxies that were missed by the COLD
GASS sample, and such galaxies can only be studied when samples are extended
down to low $M_{\rm *}$ regime. On the other hand, the lower panel of Figure.\,\ref{fig:hist2} shows that, although the three samples have different
$M_{\rm *}$ distributions, they appear to show similar $M_{\rm H\,{\sc I}}$ distributions. As a
consequence, our SMT sample and the AMIGA-CO sample span a wider $M_{\rm H_2}$ range
than the COLD GASS did, and such difference will play a role in our analysis of
the $M_{\rm H_2}$/$M_{\rm H\,{\sc I}}$ ratios (Section \ref{sec:ratio}).

\subsubsection{WISE color diagram}

The {\it WISE} all-sky near-to-middle infrared catalog with high sensitivity provides
large and homogenous samples for exploring the relationship between gas
(H$_2$ and H\,{\sc i}) and infrared emission of galaxies. Here we compare the infrared
properties derived from the {\it WISE} catalog of the three samples.

\begin{figure}[h]
  \begin{center}
	\includegraphics[scale=.4, angle=0]{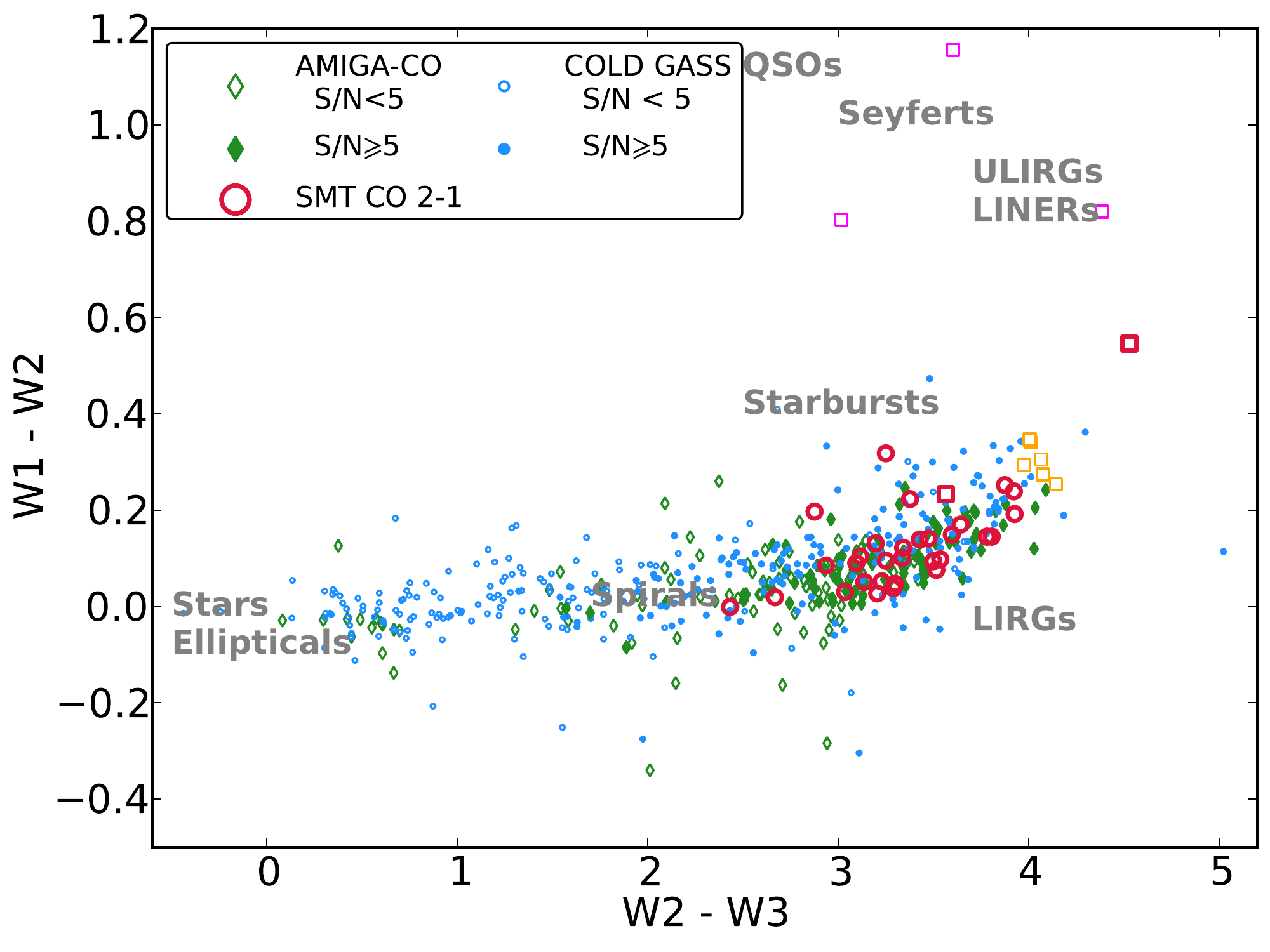}
  \end{center}
  \caption{W1$-$W2 (3.4\,$\mu$m $-$4.6\,$\mu$m) vs. W2$-$W3 (4.6\,$\mu$m $-$ 12\,$\mu$m)
  color-color diagram. Red big circles and
  squares represent the SMT sample (squares are (U)LIRGs), green diamonds and
  blue dots represent the AMIGA-CO and COLD GASS samples, respectively.
  Empty diamonds and dots show those galaxies without secure detections.
  Orange squares are COLD GASS (U)LIRGs, and magenta squares are AGNs
  classified based on the criteria W1$-$W2 $>$ 0.8. Text annotations inside
  are {\it WISE} sources classification \citep{Wright:2010}.}
  \label{fig:type}
\end{figure}

In Figure.\,\ref{fig:type} we plot the color-color diagram (W1 $-$ W2 as a
function of W2 $-$ W3) for the whole combined sample, including those sources
with only tentative or non-detection of CO. With the annotation of different
kinds of sources on the diagram \cite[see][]{Wright:2010}, it is obvious that
those galaxies without significant CO detection are more likely ellipticals,
and they tend to have `bluer' W2 $-$ W3 colors (W2 $-$ W3 $\lesssim$ 2). In
contrast, for galaxies detected in CO, there is a good consistency in {\it WISE}
colors of massive galaxies and intermediate-$M_{\rm *}$ galaxies, that they confined in
a quite small area on the diagram. The CO detected galaxies tend to be `redder'
in W2 $-$ W3 color, and such difference resembles that in the NUV$- r$ color
(Figure.\,\ref{fig:hist1}). The W2 $-$ W3 color has been suggested to be a
useful star-forming indicator \citep{Donoso:2012}, in the context that W4 band
is much less sensitive comparing with the other bands so W4 data would not be
available for as many galaxies as the other three bands are. In fact, we found
that the detection rate of CO has good correspondence with the W3 S/N, i.e.,
for both the COLD GASS and AMIGA-CO, among those galaxies with CO S/N
$\geqslant$ 5, only two sources have W3 S/N $<$ 10. 

In the COLD GASS sample, three galaxies can be classified as AGN, based on the
{\it WISE} color criteria, W1 $-$ W2 $>$ 0.8, which has been demonstrated to be
capable of selecting strong AGN/QSOs \citep{Yan:2013}. Moreover, eight COLD
GASS galaxies were classified as ULIRGs (two of them are also AGN based on the
W1 $-$ W2 criteria), and this diagram shows that these AGN/ULIRGs have
nearly the reddest IR colors among the whole combined sample, since their IR
spectrum energy distribution (SED) peaks towards longer wavelength due to
the contribution of hot dust emission originated from AGN or intense starburst.
Because the IR emission in AGN host galaxies is severely contaminated, and
deviates from normal galaxies whose IR emission is mainly contributed by
stellar populations, we exclude these AGNs in the discussion of the relation
between molecular gas and IR luminosities.

\begin{figure*}[h]
  \begin{center}
	\includegraphics[width=18cm, angle=0]{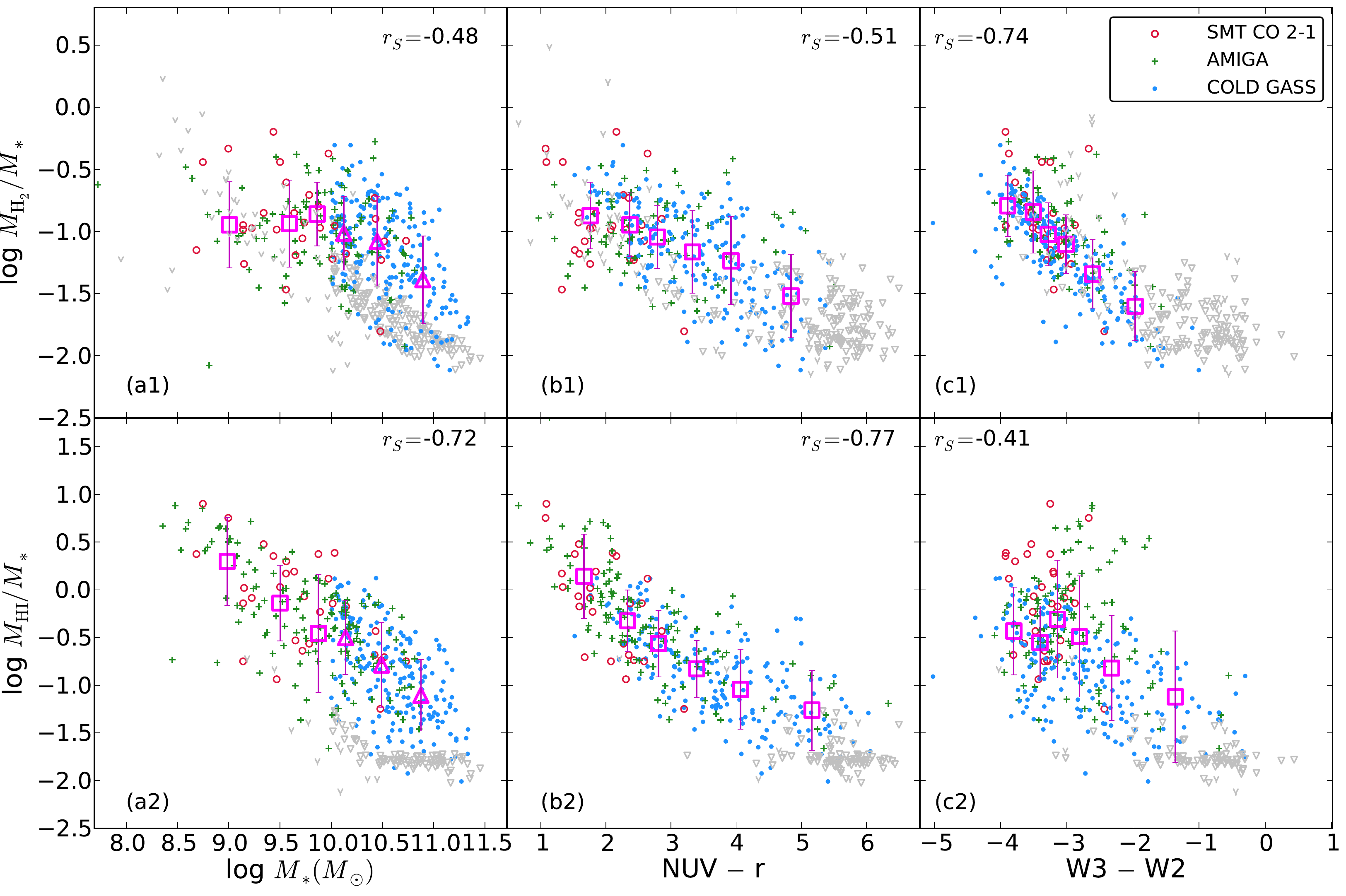}
  \end{center}
  \caption{Scaling-relations of gas fraction ($M_{\rm H_2}$/$M_{\rm *}$ ~and $M_{\rm H\,{\sc I}}$/$M_{\rm *}$) as a
  function of $M_{\rm *}$, NUV$- r$ and W3$-$W2 (12\,$\mu$m $-$4.6\,$\mu$m), respectively. Red circles, green
  crosses and blue dots represents our SMT sample, AMIGA-CO and COLD GASS,
  respectively. Dark gray triangles and light gray open arrows show gas
  fraction upper limits of COLD GASS and AMIGA, respectively. Gray solid lines
  are the linear fitting of the SMT and AMIGA-CO combined sample, with the 
  Spearman correlation coefficient $r_{\rm S1}$ shown in each panels. Blue
  dashed lines are the linear fitting for the COLD GASS sample, with the Spearman
  correlation coefficient $r_{\rm S2}$ shown in each panel. Only secure
  detections (S/N $\ge$ 5) are included in the fitting. In panel (a1) and (a2)
  the averaged log $M_{\rm H_2}$/$M_{\rm *}$ and log $M_{\rm H\,{\sc I}}$/$M_{\rm *}$ along with their scatters are
  showed with magenta squares, each bin having same number of galaxies. While
  log $M_{\rm H\,{\sc I}}$/$M_{\rm *}$ is obviously higher in lower mass galaxies, the trend of 
  increasing log $M_{\rm H_2}$/$M_{\rm *}$ with decreased $M_{\rm *}$ is very weak.}
  \label{fig:frac}
\end{figure*}

\begin{figure*}[h]
  \begin{center}
	\includegraphics[angle=0, width=18cm]{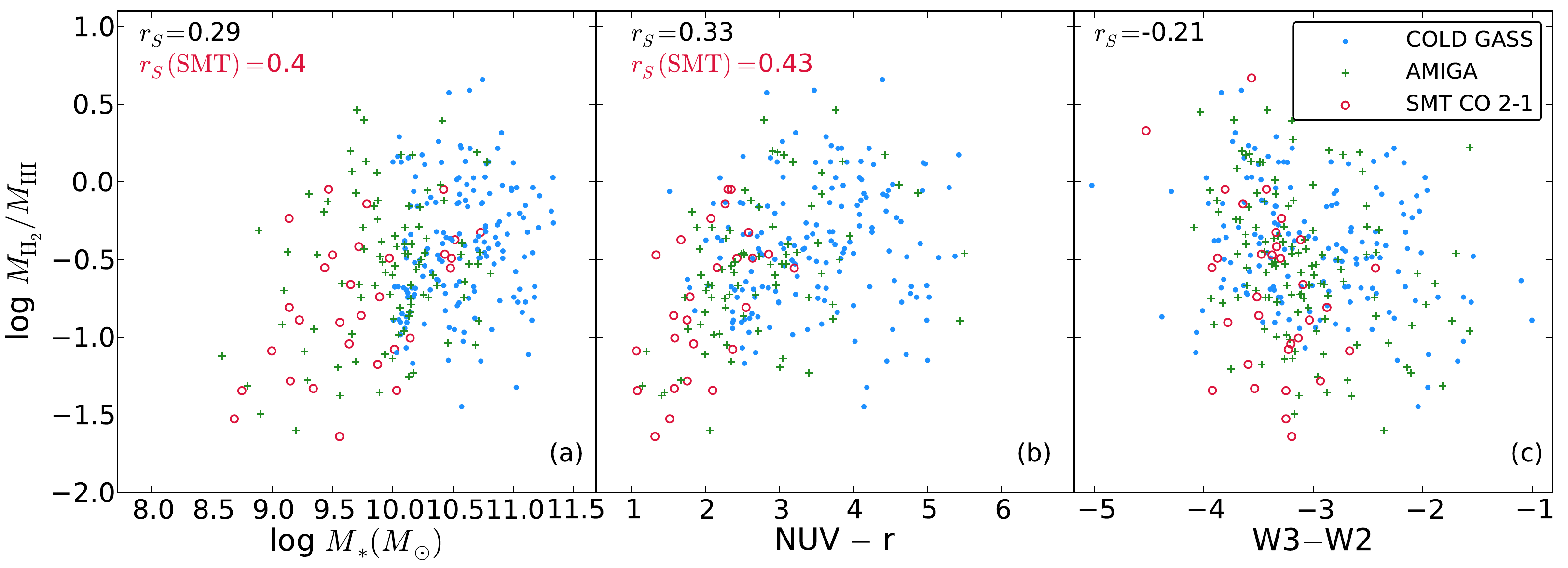}
  \end{center}
  \caption{Scaling-relations of gas ratio ($M_{\rm H_2}$/$M_{\rm H\,{\sc I}}$) as a function of 
  $M_{\rm *}$, and NUV$- r$ and W3$-$W2, respectively. Red circles, green crosses
  and blue dots represent our SMT sample, AMIGA-CO and COLD GASS, respectively.
  The Spearman correlation coefficient $r_{\rm S1}$ in each panel are for the
  SMT and AMIGA-CO combined sample, and $r_{\rm S2}$ are for the COLD GASS
  massive sample.}
  \label{fig:ratio}
\end{figure*}

\subsection{Gas fraction scaling-relations} \label{sec:fraction}

In Figure.\,\ref{fig:frac} we plot the scaling-relations between gas fraction
$M_{\rm gas}/$$M_{\rm *}$ (in this paper it refers to $f_{\rm H_2}$ $\equiv$$M_{\rm H_2}$/$M_{\rm *}$ and $f_{\rm HI}$
$\equiv $$M_{\rm H\,{\sc I}}$/$M_{\rm *}$) and several galactic properties, $M_{\rm *}$, NUV$- r$ and W3$-$W2
colors, for our SMT sample, the AMIGA-CO sample, and the COLD GASS sample.
Comparing with AMIGA and COLD GASS, our SMT observations have provided a
well selected sample which extends to lower stellar mass regime. To explore
the relationship between {\it WISE} infrared color and gas fractions, and to compare
the infrared color with NUV$- r$, we use the color W3$-$W2 so that it is
consistent with the optical color NUV$- r$, in the sense that galaxies with
greater W3 $-$ W2 values correspond to``red'', indicating the emission is
dominated by old stellar population, while ``blue'' indicating the emission is
dominated by young stars arising from on-going star formation. Note the
difference from the traditional form W2$-$W3 used in Figure.\,\ref{fig:type}
and literatures \citep[e.g.,][]{Donoso:2012}. The Spearman correlation
coefficients are showed on each panels.

In Figure.\,\ref{fig:frac}(a1) we plot $f_{\rm H_2}$ as a function of $M_{\rm *}$, and in
Figure.\,\ref{fig:frac}(a2) $f_{\rm HI}$ as a function of $M_{\rm *}$ (in log-log scale).
Combining the different samples allows the relation between $f_{\rm gas}$ and
$M_{\rm *}$ can be explored in the range between $M_{\rm *}$ $\sim 10^8$$M_{\odot}$~and $\sim
10^{11.5}$$M_{\odot}$. These two panels show two major differences between $f_{\rm H_2}$ and
$f_{\rm HI}$: First, $f_{\rm HI}$ obviously increases with decreasing $M_{\rm *}$, and in the low $M_{\rm *}$ 
end $f_{\rm HI}$ can reach as high as almost 10, while $f_{\rm H_2}$ is always lower than unity.
In the massive galaxies $f_{\rm H_2}$ significantly decreases with increasing $M_{\rm *}$, but
in low to intermediate $M_{\rm *}$ galaxies the mean $f_{\rm H_2}$ seem unchanged (see following
paragraph).  Such difference indicates that in low mass systems, most of their
gas mass are in the form of H\,{\sc i}, while their $M_{\rm H_2}$ are always less than $M_{\rm *}$.
Second, log $f_{\rm HI}$ anti-correlates much better with log $M_{\rm *}$ than log $f_{\rm H_2}$ does.
In the stellar mass range between about $10^8$ $M_{\odot}$~to $10^{11}$ $M_{\odot}$,
$f_{\rm HI}$ decreases with increasing $M_{\rm *}$ with a universal slope and the correlation
is quite prominent ($r_s = -0.72$ for the entire sample), while in Figure.\,\ref{fig:frac}(a1) the scatter is larger, and the Spearman coefficient is
lower ($r_s = -0.48$). It is intriguing to notice that the scatter of the COLD
GASS sample in panel (a1) and (a2) are quite similar, and the two figures are
mainly different in the low to intermediate $M_{\rm *}$ galaxies, which may imply that
the properties of cold gas content change quickly in galaxies of intermediate
$M_{\rm *}$, likely accompanying the transitions from H\,{\sc i} to H$_2$. 

To check the mean relations for the intermediate-$M_{\rm *}$ sample in Fig
\ref{fig:frac}(a1) and (a2), we also plot the bin averaged log $f_{\rm H_2}$ and log
$f_{\rm HI}$ (magenta squares), with nearly the same amount of sources in each $M_{\rm *}$ bin.
It is interesting that the averaged $f_{\rm H_2}$ in low mass galaxies ($M_{\rm *}$$<
10^{10}$$M_{\odot}$) seem unchanged, while the decreasing trend of the averaged
$f_{\rm HI}$ with increasing $M_{\rm *}$  is more prominent, with much smaller scatter. The
results of COLD GASS survey showed that the trend of decreased $f_{\rm H_2}$ with
increasing $M_{\rm *}$  becomes steeper at $M_{\rm *}$ $> 10^{10.5}$$M_{\odot}$~ \citep[][and
references therein]{Tacconi:2013}, and in our work the mean $f_{\rm H_2}$  at $M_{\rm *}$ $\sim
10^{10}$$M_{\odot}$~is similar to that given by the COLD GASS (7\%).  The trend of
increasing $f_{\rm H_2}$ with decreasing $M_{\rm *}$ is able to be extended down to $M_{\rm *}$ $\sim
10^{8}$$M_{\odot}$, thus we confirm that the trend becomes much shallower in lower
mass galaxies. It is interesting that our result is not only consistent with
that of the COLD GASS, but also agrees with the scenario predicted by
semi-analytical models \citep{Fu:2012}, that the $f_{\rm H_2}$ in low mass galaxies ($M_{\rm *}$
$< 10^{10.2}$$M_{\odot}$) differ not too much.

Since most of the galaxies in these plots are normal galaxies, it is natural to
see a similarity in their moderate $f_{\rm H_2}$, because high $f_{\rm H_2}$ is likely to occur
in LIRGs and/or starburst systems. Panel (a2) shows that our SMT sample
selected from the ALFALFA catalog seems to follow the same relation between
$f_{\rm HI}$ and $M_{\rm *}$ as the other galaxies, although \cite{Huang:2012} have
demonstrated that the ALFALFA population are generally more gas-rich than other
samples such as the GASS. The very high $f_{\rm HI}$ in low $M_{\rm *}$ galaxies indicates that
these systems had little integrated past star formation, and they have been
very inefficient in converting their gas into stars \citep{Huang:2012,
Blanton:2009}. Because their H\,{\sc i} is more abundant than H$_2$ (see
Figure.\,\ref{fig:ratio} and following discussions), we speculate that they
were mainly inefficient in the process of converting H\,{\sc i} to H$_2$ and forming
molecular clouds.

In Figure.\,\ref{fig:frac}(b1) and (b2) we explore how $f_{\rm H_2}$ and $f_{\rm HI}$  relate
with NUV$- r$ color, respectively. Since UV emission originates from young
stars and $r$-band emission is mainly contributed by old stars, NUV$- r$ is
strongly correlated with specific-SFR \citep[sSFR,][]{Huang:2012}, hence a good
tracer of star-forming activities. It is natural to see in panels (b1) and (b2)
that $f_{\rm H_2}$ and $f_{\rm HI}$ both depend on NUV$- r$, that $f_{\rm gas}$ is obviously
higher in blue galaxies. First of all, compared with \cite{Saintonge:2011}, we
have provided in panel (b1) more low $M_{\rm *}$ galaxies that are bluer in NUV$- r$.
Although in low mass galaxies the dust-to-gas ratio is suggested to be less due
to their lower gas-phase metallicities \citep{Blanton:2009}, we do not see any
significant difference between the samples in the relation between log $f_{\rm H_2}$ and
NUV$- r$.  With respect to the red end, \cite{Saintonge:2011} have found that
in COLD GASS galaxies the CO detection rates significantly dropped at NUV$- r$
$\gtrsim$ 5, which implied a $f_{\rm H_2}$ threshold and very little H$_2$ content in
those red and quiescent systems. Second, we show in Figure.\,\ref{fig:frac}(b2)
that in the bluest star forming galaxies, the decreasing trend of $f_{\rm HI}$ with
NUV$- r$ is steeper than that in the red galaxies, which are mostly massive.
This is consistent with the result of \cite{Catinella:2012} that H\,{\sc i}-rich
galaxies deviate from the mean relation defined by galaxies with ``normal'' H\,{\sc i}
content, and the reason for this deviation and the scatter might be the
scenario that the UV emission in red galaxies is contributed by more evolved
stellar populations, and their UV emission might not be well associated with
H\,{\sc i}. Another reason of the non-linearity was some of the galaxies with very low
gas fractions were missed in the GASS survey \citep{Catinella:2012}, and panel
(b2) shows that our sample provides more galaxies with higher
$f_{\rm HI}$, thus the non-linearity is more obvious. Third, it is interesting that log
$f_{\rm HI}$ correlate with NUV$- r$ better than log $f_{\rm H_2}$ does. It is probably because
H$_2$ resides in the central region of galaxies, where UV emission is easily
absorbed due to dense dust content, while on the other hand H\,{\sc i} dominates the
outskirt of galactic disks, where most of the UV emission can be observed
\citep{Saintonge:2011}. 

To address this issue, in Figure.\,\ref{fig:frac}(c1) and (c2) we plot $f_{\rm H_2}$ and
$f_{\rm HI}$ as a function of {\it WISE} infrared color W3$-$W2, which is another color
parameter but it is much less affected by dust attenuation than NUV$- r$ or
other optical colors. The W2 (4.6\,$\mu$m) near-infrared band is a good tracer of
stellar mass, and W3 (12\,$\mu$m) mid-infrared band covers the 11.3\,$\mu$m PAH and
can also be contributed by the dust continuum emission at 12\,$\mu$m. NUV$- r$
has been widely used to indicate galactic star formation activities and
classify galaxies into blue cloud or red sequence, and W2$-$W3 color has also
been suggested to be a good star formation indicator \citep{Donoso:2012}. Here
we compare their relations with $f_{\rm H_2}$ and $f_{\rm HI}$. Panel (c1) shows a clear
correlation between $f_{\rm H_2}$ and W3$-$W2, and the dependence of $f_{\rm H_2}$ on W3$-$W2
($r_s = -0.74$) is much stronger than that on NUV$- r$.  Panel (c1) also shows
that, most of the sources without positive CO detection are red in W3$-$W2, and
none of the galaxies with W3$-$W2 $> -1$ were detected. These ``red'' galaxies
seem to lie above the relation defined by those ``blue'' galaxies, which might
indicate that the $f_{\rm H_2}$ upper limits of the red galaxies were probably
overestimated, which is consistent with that suggested by
\cite{Saintonge:2011}, where by stacking the undetected spectra they found a
mean $f_{\rm H_2}$ less than 0.16\%. Figure.\,\ref{fig:frac}(c1) allows us to
confirm that the scatter between $f_{\rm H_2}$ and NUV$- r$ is mainly caused by dust
attenuation, and by introducing the new infrared color W3$-$W2, we show that
$f_{\rm H_2}$ has a better correlation with W3$-$W2. Since the sample used in this work
are mostly star-forming galaxies, their 12\,$\mu$m emission is contributed by
both PAH and dust continuum which originates from current star forming
activities, while their W2 (4.6\,$\mu$m) emission is mainly contributed by
stellar components. In addition, the mass of their star-forming gas (dense
molecular gas) should be proportional to the overall gas content traced by CO,
therefore we can see this strong correlation between $f_{\rm H_2}$ and W3$-$W2.

Then in Figure.\,\ref{fig:frac}(c2) we plot log $f_{\rm HI}$ as a function of W3$-$W2,
and the result shows that, the COLD GASS sample appears to be distinct from the
other two samples, and only in the COLD GASS sample log $f_{\rm HI}$  has a weak
dependence on W3$-$W2, whereas the scatter in the other two samples is much
larger.  There are some galaxies which tend to be redder than the mean relation
(magenta squares), and comparing with panel (a2) it is obvious that those
offset sources are mainly low mass galaxies. This may indicates the large
diversity of star formation efficiencies in these low $M_{\rm *}$ systems. On the one
hand, as already mentioned above, there is a large amount of H\,{\sc i} in some of the
low $M_{\rm *}$ systems but they have been inefficient in converting their H\,{\sc i} into
H$_2$ and stars, and since infrared emission originates from dust which
mixes with both atomic and molecular gas \citep{Bohlin:1978} , we do not see a
good correlation between $f_{\rm HI}$ and W3$-$W2 in low $M_{\rm *}$ galaxies. On the other
hand, in massive galaxies, they have more consistent star formation
efficiencies and consume the gas constantly thus we see a correlation between
$f_{\rm HI}$ and infrared color. Another possible explanation for the deviation of
those high $f_{\rm HI}$ galaxies in panel (c2) is that in low $M_{\rm *}$ galaxies there might
be little PAH emission due to their low metallicities, and comparing with
massive galaxies more abundant in PAH, W3 emission of low mass systems would be
contributed less by the 11.3\,$\mu$m PAH, so they are prone to be redder.

In summary, the correlations between $f_{\rm gas}$ and the two colors, NUV$-
r$ and W3$-$W2, are consistent with a galaxy evolution picture that low mass
galaxies are more gas rich and especially abundant in H\,{\sc i}, but they are
inefficient in converting their H\,{\sc i} into H$_2$ molecular clouds then stars; in
massive galaxies the star forming efficiency is more consistent and there is
less molecular gas since it has been consumed in forming stars, while the
overall decreased gas amount causes a drop in the overall gas density, and
reduce the rate of transforming H\,{\sc i} into H$_2$, with certain amount of gas can remains
in the form of H\,{\sc i}, and some of the H\,{\sc i} can be either replenished or accreted
from the environment recently, if not expelled by the interaction between
nearby galaxies.

\subsection{H$_2$-to-H\,{\sc i} gas ratio} \label{sec:ratio}

In the previous section we discuss how gas fractions $f_{\rm H_2}$ and $f_{\rm HI}$ depend on
galactic properties, respectively. It is also worth exploring the scaling
relations between $M_{\rm H_2}$/$M_{\rm H\,{\sc I}}$ and global physical parameters, because the
relationship between H$_2$ and H\,{\sc i} in galaxies is a fundamental question, yet how
this relation is affected or regulated by other galactic properties is still
unclear. As already discussed in \cite{Young:1991} and recently in the AMIGA
survey \citep{Lisenfeld:2011}, the ratio of $M_{\rm H_2}$/$M_{\rm H\,{\sc I}}$ declines in late type
galaxies, because of more H\,{\sc i} content in late type galaxies than early type
(such as S0, Sa Hubble type). \cite{Saintonge:2011} also found that log
$M_{\rm H_2}$/$M_{\rm H\,{\sc I}}$ decreases with increasing $R_{90}/R_{50}$, but the $M_{\rm H_2}$/$M_{\rm H\,{\sc I}}$ ratio has
a quite large scatter in different galaxies, although $M_{\rm H_2}$ is proportional to
$M_{\rm H\,{\sc I}}$ as expected \citep{Saintonge:2011}. 

In Figure.\,\ref{fig:ratio} we plot the scaling-relations between gas ratio
(log $M_{\rm H_2}$/$M_{\rm H\,{\sc I}}$) and the same parameters used in Figure.\,\ref{fig:frac}, $M_{\rm *}$,
NUV$- r$ and W3$-$W2. The plots show that the dependence of log $M_{\rm H_2}$/$M_{\rm H\,{\sc I}}$ on
these parameters are quite weak, and in the COLD GASS massive sample the
scatter is large, which were already found in \cite{Saintonge:2011} and can be
explained by the large scatter of log $f_{\rm H_2}$ and log $f_{\rm HI}$ in
Figure.\,\ref{fig:frac}. However, it is interesting to see in panels (a) and
(b) that for our SMT galaxies, log $M_{\rm H_2}$/$M_{\rm H\,{\sc I}}$ appears to correlate with log
$M_{\rm *}$ ($r_s = 0.4$) and NUV$- r$ ($r_s = 0.43)$, although the
color of the SMT sample tend to be blue (NUV$- r$ $<$ 4).  The $M_{\rm H_2}$/$M_{\rm H\,{\sc I}}$ ratio tend to be higher
in more massive and/or relatively redder galaxies of the SMT sample. We do not
find any significant correlation between log $M_{\rm H_2}$/$M_{\rm H\,{\sc I}}$ and W3$-$W2, which should be
resulted from the large scatter and the non-linear relation between $f_{\rm HI}$ and
W3$-$W2. The $M_{\rm H_2}$/$M_{\rm H\,{\sc I}}$ in the SMT sample are in the range of about 0.02 -- 4.7,
and only in the two LIRGs their $M_{\rm H_2}$/$M_{\rm H\,{\sc I}}$ are greater than unity, indicating
more H$_2$ than H\,{\sc i}, but in all the other normal galaxies the $M_{\rm H\,{\sc I}}$ are heavier
than $M_{\rm H_2}$. The trend in Figure.\,\ref{fig:ratio}a confirmed what was suggested
in the SINGS survey that $M_{\rm H_2}$/$M_{\rm H\,{\sc I}}$ declined at lower $M_{\rm *}$ \citep{Blanton:2009}.
And $M_{\rm H_2}$/$M_{\rm H\,{\sc I}}$ seems to increase in moderate NUV$- r$ color, is due to the
decreasing $f_{\rm HI}$ in red galaxies that is shown in Figure.\,\ref{fig:frac}(a2).

Consider Figure.\,\ref{fig:ratio} with Figure.\,\ref{fig:frac} together, we speculate
that, $M_{\rm H_2}$/$M_{\rm H\,{\sc I}}$ increases as increasing $M_{\rm *}$ in lower $M_{\rm *}$ galaxies, is an
implication of the dependence of H\,{\sc i}-to-H$_2$ transition on the $M_{\rm *}$ cumulation in
galaxies.  In the history of galaxy evolution, consuming gas to form stars is
accompanied by the accretion of surrounding diffuse gas which is mainly in
atomic form (H\,{\sc i}), and star-formation is also accompanied by the transformation
from H\,{\sc i} into H$_2$ \citep{Krumholz:2013}. On the early (young) stage, there was
enough H\,{\sc i} to be converted into H$_2$ then molecular clouds, wherein star formed
and the $M_{\rm H_2}$/$M_{\rm *}$ fraction remained at a certain level (intermediate--$M_{\rm *}$ in
Figure.\,\ref{fig:frac}(a1)); on a later stage, when a galaxy became more massive
(in $M_{\rm *}$) its sSFR (NUV$- r$) increase, then its H\,{\sc i} probably become deficient
(Figure.\,\ref{fig:frac}(a2)) as most of the H\,{\sc i} were transformed into molecular
clouds, and then the stellar mass kept growing but molecular gas fraction would
decrease (massive galaxies in Figure.\,\ref{fig:frac}(a1)). Despite the large
scatter in our plots, this scenario of the evolution of H\,{\sc i}, H$_2$ and $M_{\rm *}$, is a
possible explanation of the plausible turnover of $f_{\rm H_2}$ at $M_{\rm *}$ $\sim
10^{10}$$M_{\odot}$.  We should note two major sources of the large scatters in these
relations, one is the different star forming history and environment of
different galaxies, and the other is the larger uncertainty of $\alpha_{\rm CO}$, especially
in different kinds of galaxies, which brings in large scatter in the conversion
from CO flux to $M_{\rm H_2}$.  while COLD GASS used a different $\alpha_{\rm CO}$ for the 12 ULIRGs
(1.0 $M_{\odot}$(K\,km\,s$^{-1}$\,pc$^2$)$^{-1}$), the $M_{\rm H_2}$ of all other galaxies in the
combined sample were calculated based on a universal $\alpha_{\rm CO}$. We do not discuss
the details of the calibration of $\alpha_{\rm CO}$ in different kinds of galaxies, such as
nearby normal galaxies, dwarfs, or LIRGs, as other studies have been focusing
on this topic \citep[e.g., see][]{Bolatto:2013, Sandstrom:2013}. Therefore the
large scatter we see in these scaling-relations is largely intrinsic thus might
not be reduced easily at this stage.


\begin{figure}[h]
  \begin{center}
	\includegraphics[angle=0,scale=.5]{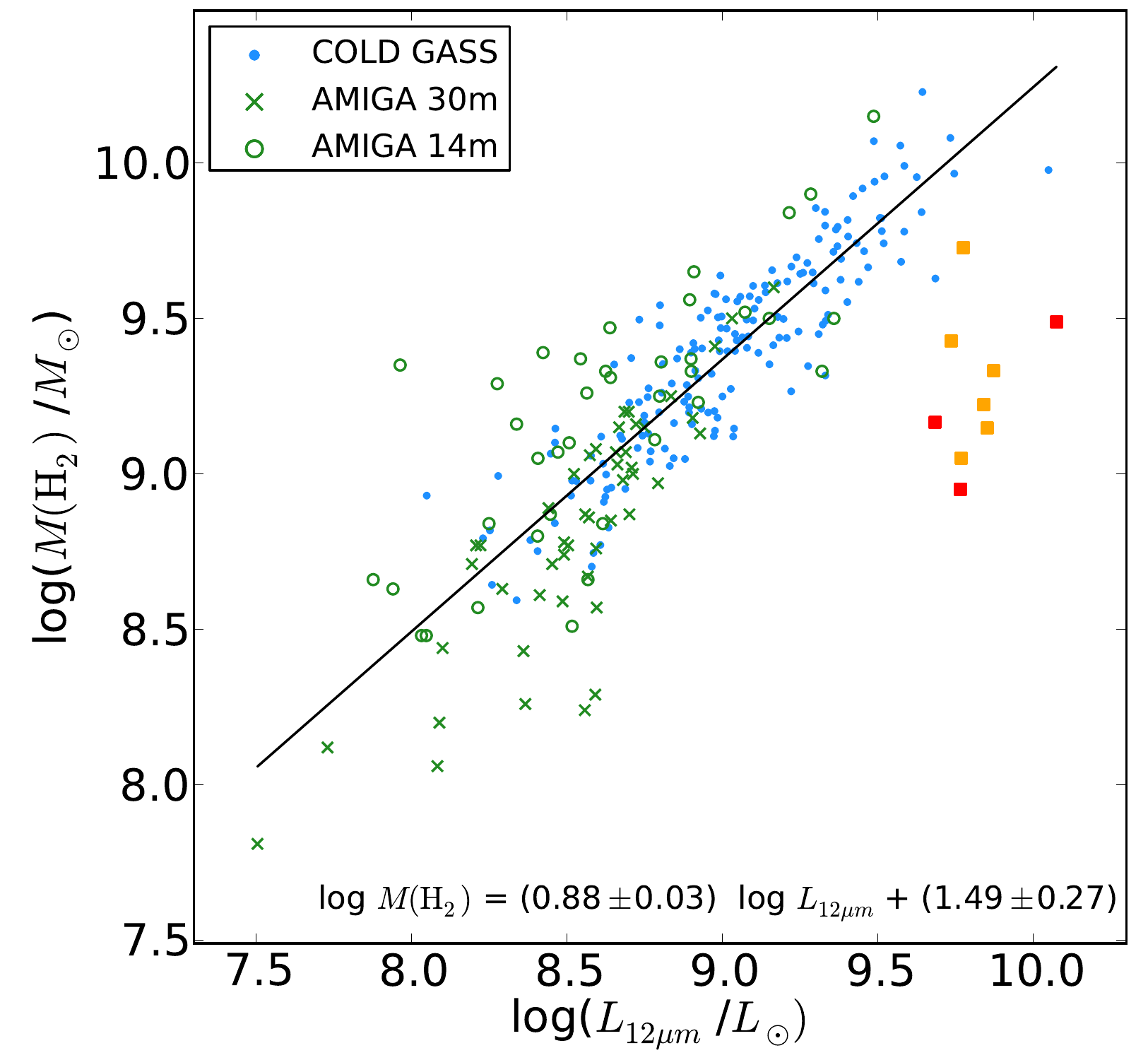}
	\includegraphics[angle=0,scale=.5]{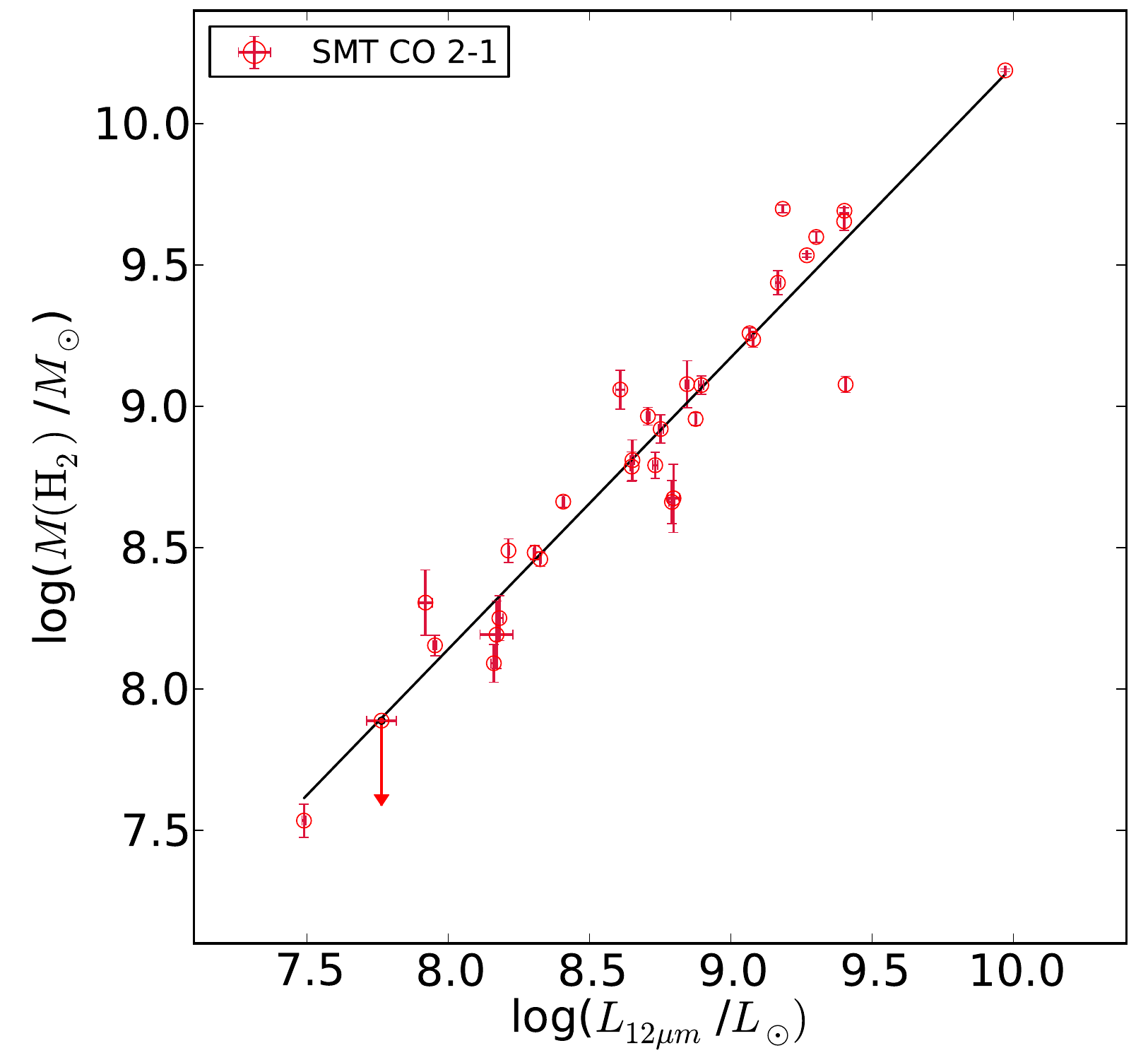}
  \end{center}
  \caption{Upper panel shows the relation between log $M_{\rm H_2}$ and $L_{\rm 12\,\mu m}$  for the 
  CO 1-0 measurement from the AMIGA sample and COLD GASS sample. Blue dots
  denote the COLD GASS, green
  crosses denote the AMIGA sample which were observed with 30-m telescope, and
  green circles denote those observed with 14-m telescope. Orange squares are
  COLD GASS (U)LIRGs and red squares are AGNs, and they are excluded from the
  fitting, which is shown as the black solid line. Lower panel shows the same
  plot for the SMT CO 2-1 sample.}
  \label{fig:cow3}
\end{figure}

\subsection{The relationship between $M_{\rm H_2}$ and 12\,$\mu$m} \label{sec:correlation}

The tight anti-correlation shown in Figure\,\ref{fig:frac}(c1) implies that
$M_{\rm H_2}$ might be well correlated with 12\,$\mu$m emission, since
$M_{\rm *}$ can be traced by the 4.6\,$\mu$m (W2). In this section we discuss the
relation between molecular gas mass $M_{\rm H_2}$, and the {\it WISE} 12\,$\mu$m emission.
\cite{Donoso:2012} has found that 12\,$\mu$m is a good tracer of star formation
especially in young galaxies, so relate 12\,$\mu$m  with $M_{\rm H_2}$ is essentially the
same question as the traditional Kennicutt-Schumit law (K-S law), which related
the SFR surface density ($\Sigma_{\rm SFR}$) and gas surface density
($\Sigma_{\rm gas}$) and has been one of the most important topics in
astrophysics \citep{Kennicutt:1998, Schmidt:1963, Kennicutt:2012}. We will not
go into details of this complicate topic in this paper, instead we only discuss
how $L_{\rm 12\,\mu m}$ relates with $M_{\rm H_2}$, and to what extend can we use $L_{\rm 12\,\mu m}$ to trace $M_{\rm H_2}$,
at least for a certain type of galaxies. Although the W4 band (22 \,$\mu$m) of
{\it WISE} suffers from relatively poor sensitivity and sparse resolution, the W3
band data still have a potential for the study of SF in galaxies.

In Figure.\,\ref{fig:cow3} molecular masses $M_{\rm H_2}$ are plotted as a function of W3
(12\,$\mu$m) luminosities $L_{\rm 12\,\mu m}$. Including in panel (a) only those galaxies with
high S/N in both CO 1-0 detection and W3 magnitude (S/N $\geqslant5$), we find that
log $M_{\rm H_2}$ is strongly correlated with log $L_{\rm 12\,\mu m}$  in all the samples. There are A
few AGNs and (U)LIRGs of the COLD GASS sample (red and orange squares in panel
a) significantly deviate from the mean relation, which is probably because
their near-to-mid infrared emission is enhanced by the warmer dust temperature
due to AGNs and/or intense starburst. We excluded these AGNs and (U)LIRGs in
the fitting, and panel (a) shows that the COLD GASS sample has a smaller
scatter comparing with the AMIGA-CO. This might be a result from the difference 
in the angular sizes of galaxies of the two samples. Since the COLD GASS
limited the galaxy distance to be larger than 100 Mpc, galaxies that are 
too large for the observations are effectively excluded, and aperture effect
did not induce significant scatter in this plot, although some of
the galaxies might be still larger than both the IRAM 30-m beam and the {\it WISE}
aperture.
On the other hand, the AMIGA sample suffers from the aperture effect because of
the smaller distances of galaxies and different instruments used in the CO
observations. First, the AMIGA included very nearby galaxies as close as a few
Mpc, and since the {\it WISE} W3 band photometry was performed with finite apertures
(the maximal aperture adopted is $\sim$ 25$\arcsec$), the data for those very
nearby galaxies only covered a small portion of the galaxy hence larger
uncertainties in both their CO and 12\,$\mu$m fluxes which were very likely
underestimated. Second, the AMIGA compiled the CO sample with data from
different telescopes with different HPBW (21$\arcsec$ for the IRAM 30-m, and
45$\arcsec$ for the FCRAO 14-m, respectively), and the W3 photometry aperture
is much smaller than the HPBW of the 14-m telescope so the W3 data were
actually collected from smaller spatial regions than that of the CO data thus
W3 emission was very likely underestimated. 

In panel (a) we plot the AMIGA galaxies with different symbols according to the
telescopes they were observed with, IRAM 30-m (green crosses) or FCRAO 14-m
(green circles). And the plot shows that the aperture effect causes most of the
AMIGA galaxies that were observed with the 14-m telescope to be weaker in $L_{\rm 12\,\mu m}$,
because the relatively small W3 aperture underestimated their $L_{\rm 12\,\mu m}$, and some
AMIGA galaxies which were observed with the 30-m galaxies tend to have smaller
$M_{\rm H_2}$ because the small beam of the 30-m telescope underestimated their CO
fluxes. These two aspects of the AMIGA-CO sample together brought large scatter
in panel (a), but the overall correlation between log $M_{\rm H_2}$ (derived from CO 1-0)
and log $L_{\rm 12\,\mu m}$ is still quite strong (Spearman coefficient $r_s = 0.90$). A least
square linear fit for Figure.\,\ref{fig:cow3}a yields: \begin{equation}
  \label{} \textrm{log}~M_{\rm H_2}({\rm CO_{1-0}})= (0.88 \pm
  0.03)~\textrm{log}~ L_{\rm 12\mu m} + (1.49 \pm 0.27) \end{equation} with a
  correlation coefficient $r=0.88$.

Figure.\,\ref{fig:cow3}b shows the same plot for the SMT sample, whose $M_{\rm H_2}$ were
converted from the CO 2-1 data. The correlation turnes out to
be even stronger than that in panel (a). Although the sample size is 
smaller than the CO 1-0 sample shown in panel (a), our sample selection has
excluded those galaxies that are interacting systems or too extended for the
observation, and the result shows that such criterion effectively alleviate the
aperture effect (see the Appendix for more discussion). 
A least square linear fit for Figure.\,\ref{fig:cow3}b yields: 
\begin{equation} \label{}
  \textrm{log}~M_{\rm H_2}({\rm CO_{2-1}})= (1.03 \pm 0.06)~\textrm{log}~
  L_{\rm 12\mu m} + (-0.12 \pm 0.48)
\end{equation}
with a correlation coefficient $r=0.96$ ($r_s$ = 0.94). It is intriguing that
the slope is very close to unity, suggesting that the $M_{\rm H_2}$ traced by CO 2-1 
is nearly proportional to $L_{\rm 12\,\mu m}$.
This slope is slightly larger than that derived from CO 1-0 in panel (a), which
can be contributed by two effects. The first effect might be the influence of
the aperture effect, as mentioned above some galaxies in the CO 1-0 sample in
panel (a) were too extended and the measurements of their $M_{\rm H_2}$ and/or $L_{\rm 12\,\mu m}$ has
large uncertainties, which might affect the fitted slope, while the SMT CO 2-1
sample suffer much less from the aperture effect since they were carefully
selected. The second effect is probably the difference between the two CO
transitions. Observations in nearby galaxies have provided a mean CO 2-1 to CO
1-0 line ratio R21 = 0.7 \citep{Leroy:2013}, and CO 2-1 has been demonstrated to
be a reliable $M_{\rm H_2}$ tracer. However, the excitation condition in different
galaxies would induce scatter in R21, especially in galaxies with different
star forming activities. In more active star-forming galaxies, the H$_2$ is more
abundant and the gas temperature is higher so CO line would tend to be
saturated hence a higher R21, while in quiescent galaxies if the temperature is
very low ($\sim$ 5K) the CO 2-1 would be less excited and R21 is lower, thus
their $M_{\rm H_2}$ is possibly underestimated with CO 2-1. Therefor this effect can
cause the slope in Figure.\,\ref{fig:cow3}b to be higher, and we consider this the
major influence.

This correlation resembles the K-S law, in which SFR can be calculated using
total infrared luminosity $L_{\rm IR}$, which is usually integrated from
12\,$\mu$m to 100\,$\mu$m IRAS data. So the correlation between $M_{\rm H_2}$ and $L_{\rm 12\,\mu m}$  is
anticipated, since in star-forming galaxies their infrared spectral energy is
dominated by young stellar populations \citep{Donoso:2012}.  Moreover, the W3
band is very sensitive to polycyclic aromatic hydrocarbon (PAH) emission at
11.3\,$\mu$m and amorphous silicate absorption (10\,$\mu$m) in nearby star-forming
galaxies \citep{Wright:2010}. The 11.3\,$\mu$m PAH emission was found to be
spatially well correlated with CO emission in high resolution observation
towards nearby star-forming galaxies \citep{Wilson:2000}, and Figure.\,\ref{fig:cow3} shows a consistent relation on the global scale.
\cite{Clemens:2013} also suggested that the dust mass in a galaxy is
proportional to its ISM content in general. One might expect that the {\it WISE} W4
band would improve this relation as 24\,$\mu$m should trace star formation better
than other near- or mid-infrared bands, but substituting W4 for W3 in Figure.\,\ref{fig:cow3}b does not improve the correlation, and shows larger scatter. We
believe this is due to the fact that comparing to the other three bands, {\it WISE}
W4 band suffers from much poorer sensitivity, more sparse resolution and
sometimes saturated images, thus for the study of extragalactic subjects that
require sensitive and accurate photometric measurements, W4 is not as useful as
the other three bands.

The significance of this correlation between 12\,$\mu$m emission and $M_{\rm H_2}$ is that
it indicates the potential application of W3 data in the prediction of
molecular gas content of galaxies, which would be useful considering that {\it WISE}
provides all-sky catalog for galaxies and it is much more sensitive than the
IRAS, whose infrared catalog has been widely used in observations to predict CO
flux. Although the ALFALFA catalog will have H\,{\sc i} data for up to $\sim$30,000
galaxies, large CO surveys to-day such as the COLD GASS have only measured
hundreds of galaxies because of current technical restrictions, and CO surveys
towards galaxies are still needed for investigating the molecular gas scaling
relations. {\it WISE} W3 data can provide CO flux estimation for observations, 
and can even provide $M_{\rm H_2}$ prediction for the sake of
statistics. Our results demonstrate that such prediction should be reliable
at least for nearby star-forming galaxies selected from the ALFALFA-{\it WISE}
matched sample.  It should be noted that the sample used in our discussion are
mainly star-forming galaxies, and careful treatment is still needed to verify
whether this method of using $L_{\rm 12\,\mu m}$ to trace $M_{\rm H_2}$ can be applied to other kinds
of galaxies, such as early type galaxies, as 12\,$\mu$m PAH emission can also
originate from old stellar population, and in those galaxies without
significant star forming activities, the 12\,$\mu$m emission will be likely
dominated by such old stars. We also caution that PAH is not an unambiguous
tracer of star formation and such correlation might actually reflect that these 
galaxies share similar properties at 12\,$\mu$m , which is contributed by both 
dust continuum and PAH emission. It would be interesting to study the behaviors of
W3 emission in a variety of galaxy samples. Further verification of this
technique and its usage in the study of gas scaling relations in such large
sample, will be addressed in future work.

\section{Summary} \label{sec:summary}

We selected a sample of 32 gas-rich normal star-forming galaxies from the
ALFALFA-{\it WISE} matched sample, which span a stellar mass range between $\sim
10^{8-10.6}$$M_{\odot}$, and carried out CO 2-1 observations with the SMT 10-m
telescope. We got high CO 2-1 detection rate and only one galaxy without
significant CO detection (see Figure.\,\ref{fig:spec}). The calculated $M_{\rm H_2}$ and
$M_{\rm *}$ are presented in Table \ref{tab:measurements}.  

CO 1-0 data are compiled from other galaxies surveys, the COLD GASS and AMIGA,
and we use them along with our SMT sample to explore the gas scaling relations
with $f_{\rm HI}$ and $f_{\rm H_2}$. The AMIGA-CO sample has similar properties as our SMT
sample, in the sense that they share common $M_{\rm *}$, $R_{90}/R_{50}$ and NUV$- r$
distributions (Figure.\,\ref{fig:hist1}), so combining the SMT sample with the
AMIGA-CO sample allows us to improve the sample size of low- to
intermediate-$M_{\rm *}$  galaxies($M_{\rm *}$ $< 10^{10}$ $M_{\odot}$), and can be compared with the
COLD GASS massive sample ($M_{\rm *}$ $> 10^{10}$ $M_{\odot}$). Since we include a number of
galaxies of low $M_{\rm *}$, the dynamical range in plotting the relation between
$f_{\rm H_2}$ and $M_{\rm *}$ as well as other galactic properties is enhanced comparing to the
COLD GASS. Our main results include:

\begin{enumerate}

\item Bin-averaged $f_{\rm H_2}$ and $f_{\rm H_2}$ are derived for the whole sample, which shows
  that $f_{\rm HI}$ obviously increase in galaxies of lower $M_{\rm *}$, while $f_{\rm H_2}$ almost keep
  unchanged in low- to intermediate-$M_{\rm *}$  galaxies ($M_{\rm *}$ $< 10^{10}$ $M_{\odot}$). This
  result is consistent with that of the COLD GASS at similar $M_{\rm *}$ ($\sim
  10^{10}$ $M_{\odot}$), but we are able to extended it to lower $M_{\rm *}$ regime. We also
  induce a new parameter W3$-$W2 and compare the scaling relations between
  $f_{\rm gas}$ and two colors, NUV$- r$ and W3$-$W2, which are all
  star-forming indicators (Figure.\,\ref{fig:frac}).  Our results show that
  while NUV$- r$ is anti-correlated with log $f_{\rm HI}$ tighter than $f_{\rm H_2}$, W3$-$W2
  has tighter anti-correlation with log $f_{\rm H_2}$ than log $f_{\rm HI}$. 

\item The gas ratio (log $M_{\rm H_2}$/$M_{\rm H\,{\sc I}}$) scaling relations are also explored, and
  they all show large scatter (Figure.\,\ref{fig:ratio}). This can be
  attributed to the scatter in the $f_{\rm H_2}$ and/or $f_{\rm HI}$ scaling relations. Only in
  the SMT sample we see correlations between log $M_{\rm H_2}$/$M_{\rm H\,{\sc I}}$ and log $M_{\rm *}$ as well
  as NUV$- r$, that more massive and/or redder galaxies have higher $M_{\rm H_2}$/$M_{\rm H\,{\sc I}}$.
  
\item We compare the relation between $M_{\rm H_2}$ and infrared properties derived from
  the {\it WISE} catalog, and find an excellent correlation between log $M_{\rm H_2}$ and log
  $L_{\rm 12\,\mu m}$, the 12\,$\mu$m luminosity. The slope of linear fittings of our SMT CO 2-1
  sample is very close to unity (1.03$\pm$0.06), while the slope in the CO 1-0
  sample is slightly shallower (slope = 0.88$\pm$0.03).  Considering the {\it WISE}
  has provided all-sky data with high sensitive and has significant overlap
  with the ALFALFA H\,{\sc i} catalog, one can use this relation to predict CO flux
  for observations, or even estimate $M_{\rm H_2}$ for a large amount of galaxies, at
  least for the ALFALFA-{\it WISE} matched sample which are dominated by gas-rich
  star-forming galaxies. Note that this method might not be used in galaxies
  dominated by old stellar populations, since a significant part of the 12
  \,$\mu$m emission would originated from old stars thus no longer only trace
  young stars which are tightly correlated with molecular gas.

\end{enumerate}

\acknowledgements
We thank the anonymous referee for careful review and valuable suggestions,
the SMT staffs for their help during the observations, 
Ran Wang for her help with the observation,
and U. Lisenfeld for kindly making the H\,{\sc i} data available to us.
X.J. acknowledges support for this work from the Strategic Priority Research
Program ``The Emergence of Cosmological Structures’ of the Chinese Academy of
Sciences (CAS), grant XDB09000000.
This work is supported under the National Natural Science Foundation of China
(grants 11273015 and 11133001), the National Basic Research Program (973
program No. 2013CB834905), and Specialized  Research Fund for the Doctoral
Program of Higher Education (20100091110009).  
Z-Y.Z acknowledges support from
the European Research Council (ERC) in the form of Advanced Grant, {\sc
cosmicism}.
This publication makes use of data products from the Wide-field Infrared Survey
Explorer, which is a joint project of the University of California, Los
Angeles, and the Jet Propulsion Laboratory/California Institute of Technology,
funded by the National Aeronautics and Space Administration.  
The Arecibo Observatory is part of the National Astronomy and Ionosphere Center
which is operated by Cornell University under a cooperative agreement with the
National Science Foundation.  
Based on observations made with the NASA Galaxy Evolution Explorer.  GALEX is
operated for NASA by the California Institute of Technology under NASA contract
NAS5-98034.
Funding for SDSS-III has been provided by the Alfred P. Sloan Foundation, the
Participating Institutions, the National Science Foundation, and the U.S.
Department of Energy Office of Science. The SDSS-III web site is
\url{http://www.sdss3.org/}. 
SDSS-III is managed by the Astrophysical Research Consortium for the
Participating Institutions of the SDSS-III Collaboration including the
University of Arizona, the Brazilian Participation Group, Brookhaven National
Laboratory, University of Cambridge, Carnegie Mellon University, University of
Florida, the French Participation Group, the German Participation Group,
Harvard University, the Instituto de Astrofisica de Canarias, the Michigan
State/Notre Dame/JINA Participation Group, Johns Hopkins University, Lawrence
Berkeley National Laboratory, Max Planck Institute for Astrophysics, Max Planck
Institute for Extraterrestrial Physics, New Mexico State University, New York
University, Ohio State University, Pennsylvania State University, University of
Portsmouth, Princeton University, the Spanish Participation Group, University
of Tokyo, University of Utah, Vanderbilt University, University of Virginia,
University of Washington, and Yale University.


\end {CJK*}
\bibliographystyle{apj}
\bibliography{bibtex_xjiang}{}

\appendix
\section{Aperture effect} 

Our study is affected by aperture effect from both the radio telescopes and
{\it WISE}. On the one hand, the single dish telescopes (IRAM 30-m, FCRAO 14-m and
SMT 10-m) mentioned above all have a finite aperture size (HPBW) and can only
cover a portion of the CO emission from a galaxy observed if its angular size is
significantly larger than the beam size of the dish. Thus for the majority of
the sources, especially those observed with 30-m telescope whose HPBW was only
22\arcsec~at 115\,GHz, an aperture correction from the observed CO flux to
total CO flux was necessary. In the latest COLD GASS data release and the
AMIGA sample, they both assume that the radial distribution of CO and molecular
gas is exponential, and the CO scale length $r_e$ correlates well with the
optical radius at the 25mag isophote, $r_e = 0.2~r_{25}$. Thus the correction
from the observed CO flux to the total flux based on this model can be
calculated.  The technical details about the aperture correlation method can be
found in \cite{Lisenfeld:2011} and \cite{Saintonge:2012}. 

On the other hand, the standard photometry pipeline of {\it WISE} adopted a series
of circular apertures to measure the magnitudes of a galaxy, and the maximum
aperture size used is 24.75\arcsec. Thus for some large galaxies their W3 will
suffer from the same aperture issue, so the study of the relation between
molecular gas and W3 relies on improved accurate photometry. However, since the
angular sizes of the galaxies in our selected SMT sample is relatively suitable
for both {\it WISE} and the SMT beam (33\arcsec~at 230 GHz) and could be effectively
covered, we did not adopt this aperture correction for this sample, and Figure.\,\ref{fig:cow3}b already shows a very robust correlation. There is a galaxy in
Figure.\,\ref{fig:cow3}b, whose $L_{\rm 12\,\mu m}$ is about 9.4 $L_{\odot}$, showing deviation from the
fitted line, and we confirmed that its enhanced $L_{\rm 12\,\mu m}$ is due to the
contamination from a nearby star, and the confusion could not be reduced
because of the relatively large beam of {\it WISE}. Nevertheless this source does
not affect the overall fitting results even it is included in the sample for
statistics.

\section{SMT CO 2-1 Spectra} \label{sec:spectra}
\small
\begin{figure}[ht]
\begin{center}
	\includegraphics[angle=0,scale=0.85]{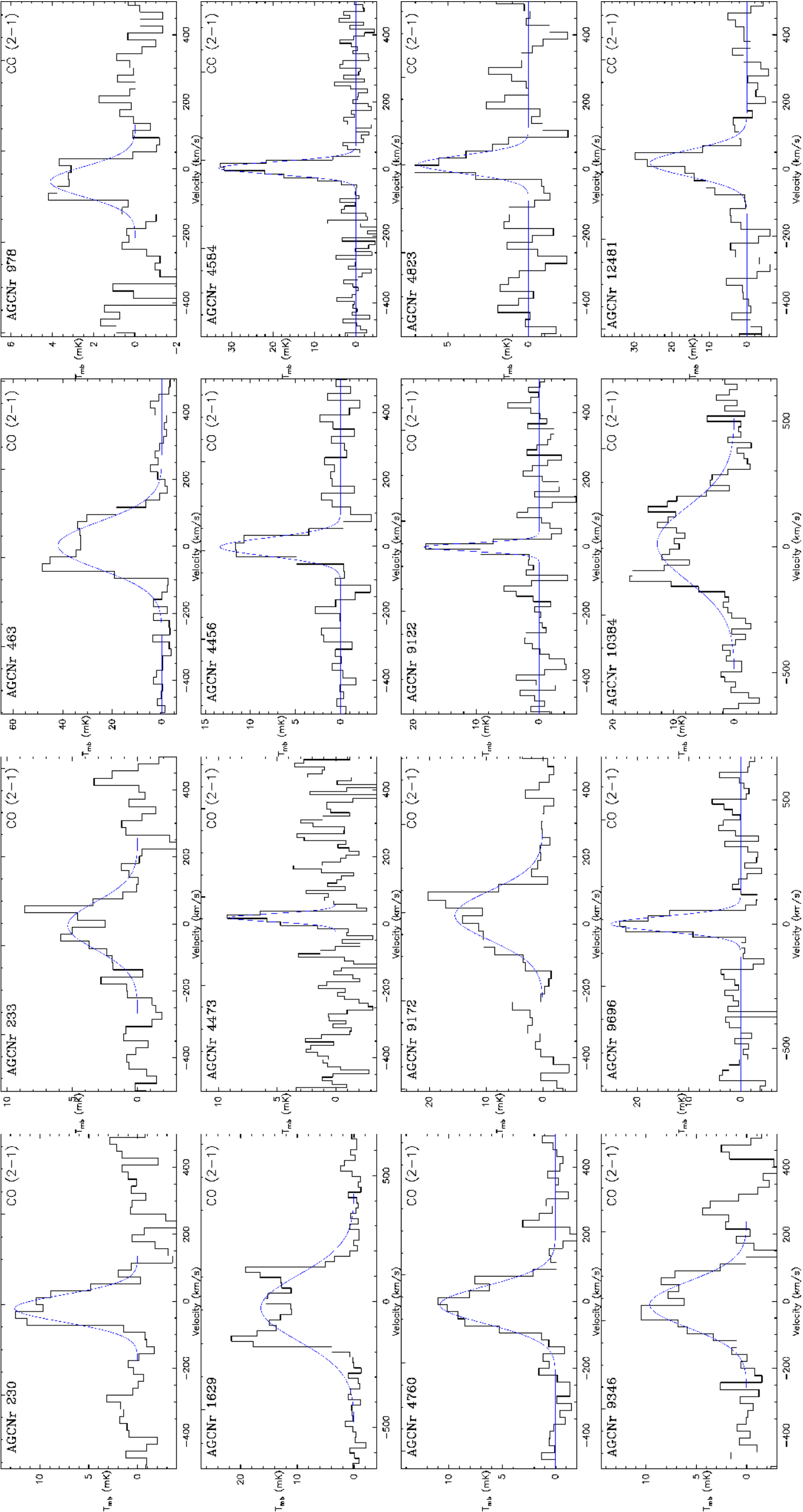}
  \caption{CO 2-1 spectra of the SMT sample (1). Source names from the 
  ALFALFA are denoted in the upper left corner of each panel. The 
  Gaussian profiles are only for reference but not used in the flux
  calculations.}
  \label{fig:spec}
\end{center}
\end{figure}

\begin{figure}[ht]
\begin{center}
	\includegraphics[angle=0,scale=0.85]{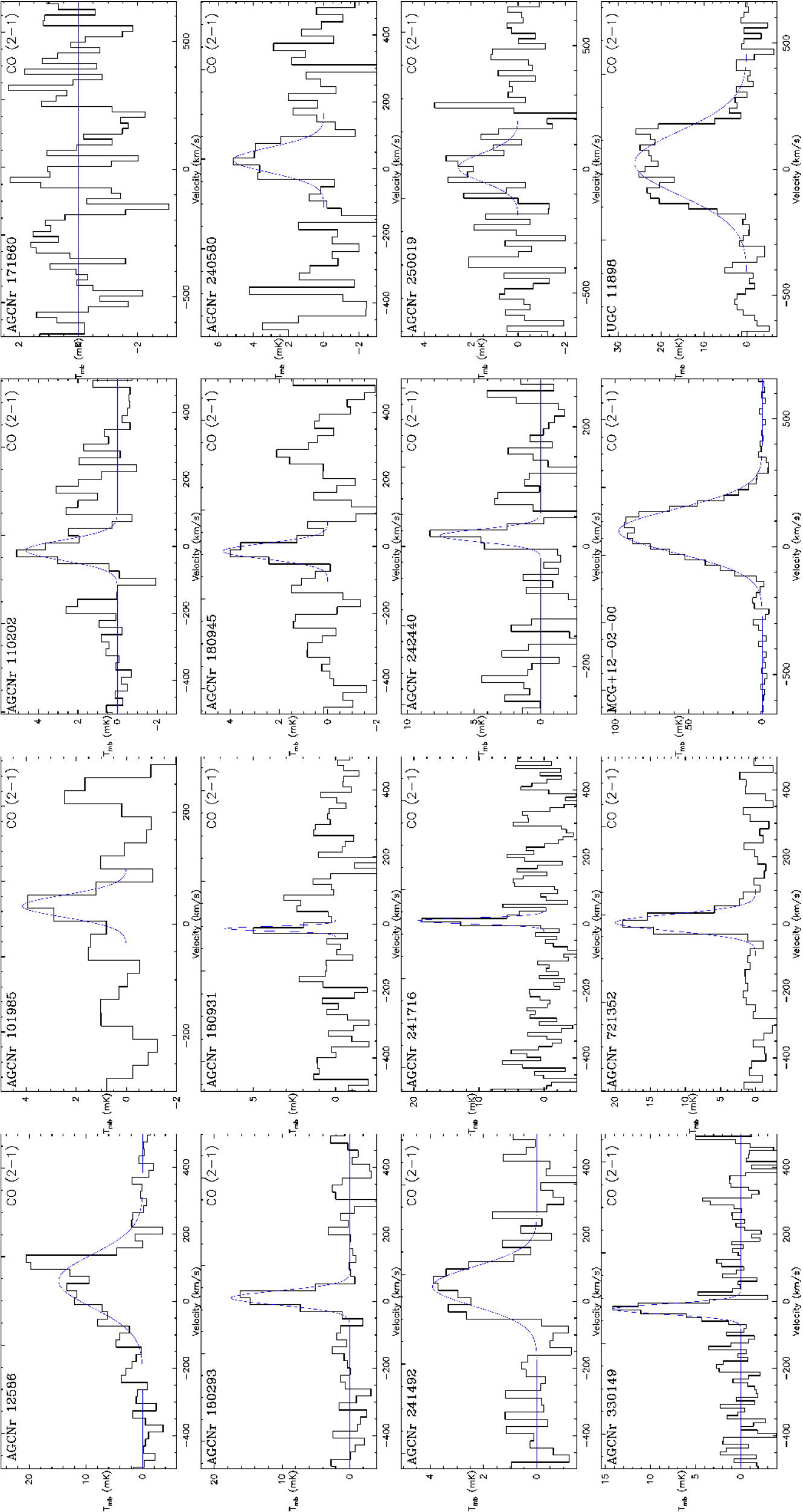}
\setcounter{figure}{6}
  \caption[]{CO 2-1 spectra of the SMT sample (2). Source names from the 
  ALFALFA are denoted in the upper left corner of each panel. The 
  Gaussian profiles are only for reference but not used in the flux
  calculations.}
\end{center}
\end{figure}


\end{document}